Work package 3

Task 3.1 Creation and experimentation usage of working environment with tools for pre-processing of digital audio and video archive objects.

# The PCI Ontology - Version 4.

Presentation and discussion of an ontology for the description and indexing of audiovisual resources in the field of cultural heritage of minorities and indigenous people.

author: Peter Stockinger

FMSH/ESCoM – Paris, 15/01/2007

ESCoM/Logos report *04-Logos-07*

# Contents

# 1/ The general context

## 1.1/ The pilot "Cultural Heritage of Minorities and Indigenous People" (P.C.I.)

In this paper, an ontology for the description and indexing of the contents of audiovisual resources will be presented. The concerned domain of reference (or: knowledge domain) is the cultural heritage of minorities and indigenous people, hence the name or title of this ontology, *PCI ontology* (standing for the French "**P**atrimoine **C**ulturel de Minorités et Peuples **I**ndigènes"). The whole audiovisual corpus can be accessed and explored on the PCI audiovisual web portal (figure 1): http://semioweb.msh-paris.fr/corpus/pci/fr/

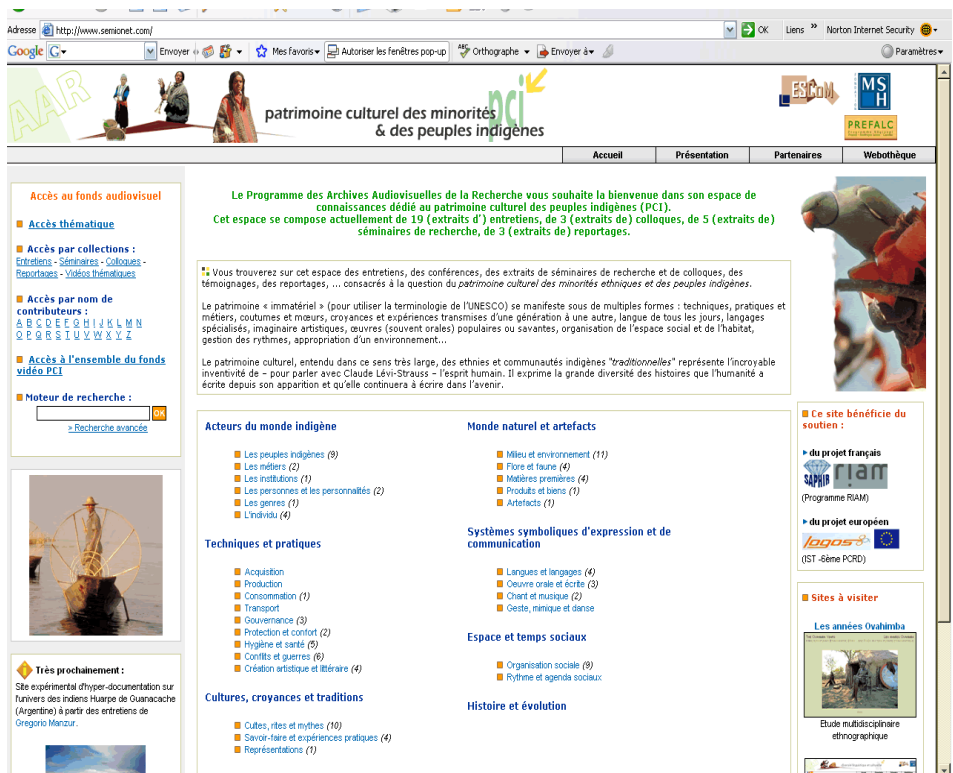

(*figure 1*: the homepage of the PCI audiovisual web portal)

The audiovisual corpus is composed of interviews with anthropologists, sociologists, archaeologists, linguists and other specialists and professionals of the concerned domain, research seminars, conferences, workshops as well as scientific film footages, documentaries films of expositions and performances, etc. It is mainly in French language but videos are also available in English, Spanish and several other languages. But as a re-purposed (re-authored) corpus during and for the Logos project, it will be available as a multilingual annotated one, i.e. as a corpus with content annotations (translations, summaries, key words, …) in English, Russian, Spanish and partially in Turkish. From a quantitative point of view, the corpus represents at the end of 2006 about 100 hours of videos. It is an open corpus, viz. a corpus which is continuously enriched by new audiovisual resources.

Very roughly speaking, the domain of reference is centrally concerned of what UNESCO calls the immaterial cultural heritage[1] of minorities and indigenous people: their cultural identity and self, their traditions and customs, their social way of life, the organisation of their daily life, their artistic and religious specificities, but also their technical know how, their practical experiences for inhabiting and controlling an ecological niche, their relationships with the social environment in which they are living as well as the often very difficult and unfortunate (social, cultural, demographic, …) evolution of them as a social group. It is through, also, that one of the missions of the PCI web portal is, to give an audience to the political and social requirements of minorities and indigenous people – in Europe and elsewhere. Finally, even if the *intangible* aspect of cultural heritage is stressed, this accent does not constitute an exclusive limitation of the PCI web portal: also issues of *material* culture, *embodied* culture, housing culture, culture of eating and drinking, and so on are dealt in the contributions the PCI source corpus

Figure 1 show the principal access and exploration possibilities of the actual version of the PCI portal homepage. One central access modality is proposed by a thematic catalogue in the central part of the PCI homepage. The themes of this catalogue fit with the *basic vocabulary* of the PCI ontology. Therefore, even if in its actual version, the catalogue is a static one, it will become a main modality of accessing and exploring the audiovisual resources once it will have been replaced by its dynamic version helping a user to localise relevant information and, then, to explore potentially relevant contexts of a localised information. This should become possible through the Logos project, viz. the indexing of audiovisual resources for multiple, user and context attuned exploitations.

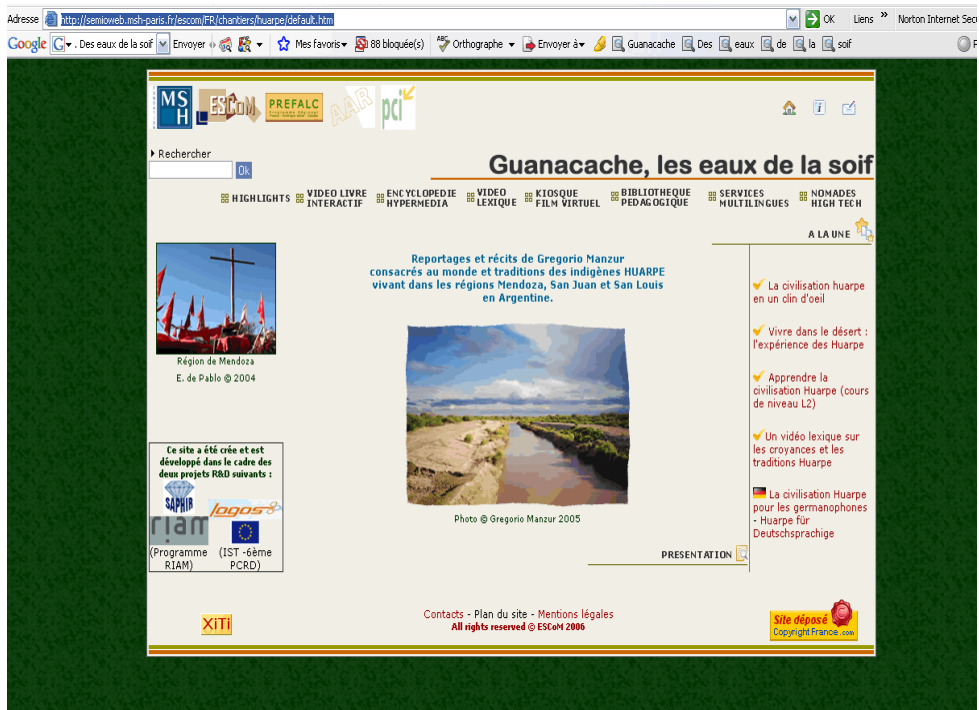

(*figure 2*: the homepage of the experimental web site dedicated to the cultural heritage of the Huarpe community with accesses to thematic folders, video-lexicons, multilingual versions, pedagogical folders, etc.)

Other exploitation modalities of an indexed audiovisual resource are experimentally shown in figure 2. Figure 2 represents the home page of an experimental web site belonging to the PCI audiovisual web

---

[1] Cf. UNESCO web site, « What is intangible cultural heritage »
(http://www.unesco.org/culture/ich_convention/index.php?lg=EN&pg=00003)

portal – an experimental web site dedicated to the cultural heritage of a rural population – the Huarpe - living in the North of Mendoza in Argentina. A given multimedia (audiovisual, visual, textual, sound,…) corpus is reused in different publishing models adapted to specific user expectations and presumed uses and exploitations: interactive virtual films, thematic folders, video-lexicon, multilingual versions, pedagogical folders, video clips for "nomads" (mobile phones, ipods, …). These services, experimented, actually, on the web site dedicated to the cultural heritage of the Huarpe community, should become available for the whole PCI web portal.

One critical milestone for progressing in this direction is the description and indexing of the PCI audiovisual resources. The description and indexing of audiovisual resources presuppose a "*description language*" – a *vocabulary* for describing them as well as description *schemas* by the means of which characteristic content features of them may be elicited.

The rest of this report is dedicated to the presentation of this "description language" – of the vocabulary of description (a meta-linguistic terminology) and of the description schemas based on the defined vocabulary and formal rules provided by the conceptual graph theory[2].

## *1.2/ Scope and objectives of the PCI ontology*

The central objective for which we conceive, realize and maintain the PCI ontology is the description of the content of discourses registered in audiovisual files that compose the resources of PCI archive.

This is a rather specific orientation which claims at least two different ontologies or, as for pragmatic reasons decided in this project, two different facets of one overall ontology:

1. the **domain ontology** (the referential facet "**World_PCI**")
2. the **discourse ontology** (the narrative facet "**Discourse Description**").

The narrative facet "Discourse Description" is necessary for enabling us to describe the discourses hold by the researchers, professionals, etc. about the domain of cultural heritage. The referential facet "World_PCI" is necessary for describing what the researchers, professionals, etc. are developing in their discourses.

Moreover, given the fact that indexed audiovisual files should become reusable resources in different publishing contexts or again for different user profiles and contexts of use and exploitation, a third ontology or, for pragmatic reasons, a third facet of one overall ontology is necessary for doing this, i.e.:

3. the **pragmatic description ontology** (the facet "**Pragmatic Description**").

This third facet will become a part of the Logos authoring studio. It figures as a specific facet within the PCI ontology (at least in the actual version 4 of the PCI ontology) in order to allow those who have to describe and index the PCI audiovisual corpus to "mark" identified and thematically described segments with respect to their interest or relevancy for a specific type of publishing genre and publishing context. Once the Logos authoring studio exists, these "marks" will be dropped.

In any case, in looking on especially to narrative fact of the PCI ontology – the Discourse Description part – it should be evident that the PCI ontology is not conceived (not priory conceived) as a sort of classification means of objects or digitised objects as this is the main objective of existing thesaurus or other terminological resources in cultural heritage such as (the, by the way, very impressive) Getty's AAT[3] and Getty's Categories of Description of Works of Art[4], the CIDOC's Guidelines for Museum Object Information[5], the British Museum Object Names Thesaurus[6], etc.

---

[2] Cf; the works of Michel Chein, David Genest, Michel Leclère, Marie-Laure Mugnier on the COGITANT homepage: http://cogitant.sourceforge.net/ and the LIRMM homepage "Représentation des connaissances et raisonnements" (R.C.R) : http://www.lirmm.fr/xml/fr/0098-03.html

[3] Cf. the web site of Getty's Art & Architecture Thesaurus:
http://www.getty.edu/research/conducting_research/vocabularies/aat/

The PCI ontology is also not (priory or exclusively) a cataloguing tool for librarian services and for which exist indeed different – more or less useful - terminologies and thesaurus having sometimes the status of content description standards. Well-known resources are, for instance, the Library of Congress' subject classification[7] or UNESCO's thesaurus of social and human sciences[8].

Much more nearer to the scope and objectives of the PCI ontology is the TEI – *Text Encoding Initiative*[9]. This – especially among linguists – well known initiative began in 1988 and aims at the establishment of guidelines for dealing with problems of text documentation, text representation, text interpretation. "Text" means written text as well as spoken text ("discourse", in a narrow acceptation) and recovers literary texts (verse, dramatic text,…), scholarly texts (dictionaries, handbooks, ..) as well as any other sort of text belonging to the daily life or professional communication. The main objective of TEI is to determine *structural* features of these types or genres of written or spoken texts. The PCI ontology integrates and reuses those TEI categories in its "Discourse Description" facet which are relevant for its type of texts broadly speaking (i.e. spoken texts belonging to the scientific communication realm: interviews, conferences)

But also with respect to the TEI, the scope of the PCI ontology is different. It is not or not only concerned in exhibiting and classifying *structural text features* of the organisation of a written or spoken text (such as a drama or a dictionary).

The central scope and objective of the PCI ontology is a *communicational* one. It is to provide the indexer with a vocabulary and schemas for:
*1/ identifying and describing information or knowledge delivered by the interviews, conferences, etc. composing the audiovisual (textual,…) corpus of the PCI portal and*
*2/ attuning this delivered information and knowledge to specific user profiles and contexts.*

This means: The PCI ontology has to recover *a/* the knowledge of the reference domain (the cultural heritage of minorities and indigenous people), *b/* the point of view of the speaker – the researcher, the educationalist, the professional working in the field of the cultural heritage of minorities and indigenous people, *c/* the point of view of those who are (potentially) interested in this domain and in the knowledge produced by the speaker. These three focuses are represented, in the PCI ontology, by the three main facets:
- "Discourse Description";
- "World_PCI Description";
- "Pragmatic Description".

Closely related to the communicational scope and objective of the PCI ontology is a R&D project leaded in the CWI of The Netherlands dedicated to the description/indexing of multimedia objects, the "speaking" about them in form of progressive narratives and the multimedia presentation of a narrative[10]. The CWI project *"Towards ontology-driven discourse: from semantic graphs to multimedia presentations"* is based on three "ontologies":

---

[4] Cf. the web edition of Getty's Description of Works of Art, edited by Murtha Baca and Patricia Harpring: http://www.getty.edu/research/conducting_research/standards/cdwa/index.html
[5] Cf. the CIDOC web site: http://www.willpowerinfo.myby.co.uk/cidoc/guide/guide.htm
[6] Cf. the web site of the BM Object Names Thesaurus: http://www.mda.org.uk/bmobj/Objintro.htm
[7] Cf. the Library of Congress Classification Outline: http://www.loc.gov/catdir/cpso/lcco/lcco.html
[8] Cf. the hierarchical list of UNESCO's Thesaurus – facet "Social and Human Sciences": http://www2.ulcc.ac.uk/unesco/4.htm
[9] Cf. the TEI web site: http://www.hti.umich.edu/t/tei/
[10] Geurts, Bocconi, van Ossenbruggen and Hardman, "Towards ontology-driven discourse: from semantic graphs to multimedia presentations" (Amsterdam, CWI, 2003); http://db.cwi.nl/rapporten/abstract.php?abstractnr=1351

1. a domain ontology (here: painters and paintings);
2. a "discourse ontology" for the progressive dealing with selected information in the domain (here: biography, portrait, ...);
3. and a multimedia presentation ontology (here: a web site).

For example, information about Rembrandt, his life and work as well as more specifically about several paintings of him are represented in form of a "semantic graph" (i.e. a thematic configuration), developed either as a *biography* of the painter, a *portrait* of the painter, an *interpretation* of a painting, etc. and presented following different "presentation styles" (i.e. interfaces that diverge from a topographical, graphical, ... point of view).

Another, very appealing and innovative example from a similar CWI project – the *Vox Populi* project[11] aiming at the automatic edition of video documentaries – is the following one: a collection of small interviews of people developing their point of view of the American war in Afghanistan against the Taliban are indexed on a rhetorical level with respect mainly to the position expressed by the people (against or in favour of …). This material is, then, developed as a rhetorical narrative (a documentary) that tries to stress a specific point of view and the backing of it (following a well known argumentation schema, the Toulmin schema[12]).

The CWI *Vox Populi* project is certainly the most relevant and most persuasive one for understanding the scope and objectives of the PCI ontology which will be used for the republishing of an audiovisual corpus dedicated to the cultural heritage of minorities and indigenous people as outlined in another ESCoM report for the Logos project[13].

Similarly to the *Vox Populi* project, the scope of the PCI ontology is to catch the knowledge for a domain of reference that is uttered by a speaker with respect to a target group/a target context of use. This specific orientation concerns mainly what is called (scientific) discourse analysis, thematic or "subject" description of digital resources or again content description – approaches and methods by the means of which the central objectives of the PCI pilot should be realised, viz.:

1. the republishing of the audiovisual corpus of the PCI web archive by the means of specific publishing genres adapted to specific user profiles and contexts of use;
2. the fine tuned localisation of (with respect to the needs or interests of a user) value added information in the audiovisual corpus and the dynamic exploration of potentially relevant contexts of a localised, value added information.

In order to close this introductory chapter, let us mention that the descriptive objectives of which the PCI ontology is concerned, has partially it's roots in text and discourse comprehension and also generation researches and, here more specifically again, in those that are related to narrative, rhetoric and semiotic inputs. One of the best known examples in this field is probably K. McKeown's[14] research on text comprehension and generation presented in a book published in the middle of the 80th, the different computational narrative grammars, the cognitive and computational theories on discourse production and exchange with questions related to the attitude, the believes, the knowledge of the speakers and hearers, and finally the rhetorical structure theory[15] (RST).

---

[11] Cf. the Vox Populi interface: http://homepages.cwi.nl/~media/demo/VoxPopuli/
[12] For a short and comprehensive introduction to Toulmin's well-known argumentation schema, cf. the University of Nebraska web site: http://www.unl.edu/speech/comm109/Toulmin/layout.htm
[13] Cf. Peter Stockinger, Specification of Logos authoring processes and publishing genres (Paris, ESCoM 2006) http://semioweb.msh-paris.fr/escom/ressources_enligne/projets_recherche/06_09_logos/WP3_4_Interview_reauthoring_process.pdf
[14] Cf. Kathleen McKeown's homepage: http://www1.cs.columbia.edu/~kathy/
[15] Cf. the RST community web site: http://www.sfu.ca/rst/

# 2/ A global presentation of the PCI ontology

## 2.1/ The general structure of the vocabulary

Figure 3 shows the overall structure of the PCI ontology realised with CoGui[16], an ontology editor within the realm of the conceptual graph theory developed by Alain Guiterrez and the team "*Représentation des Connaissances et Raisonnements*"[17] of the LIRMM.

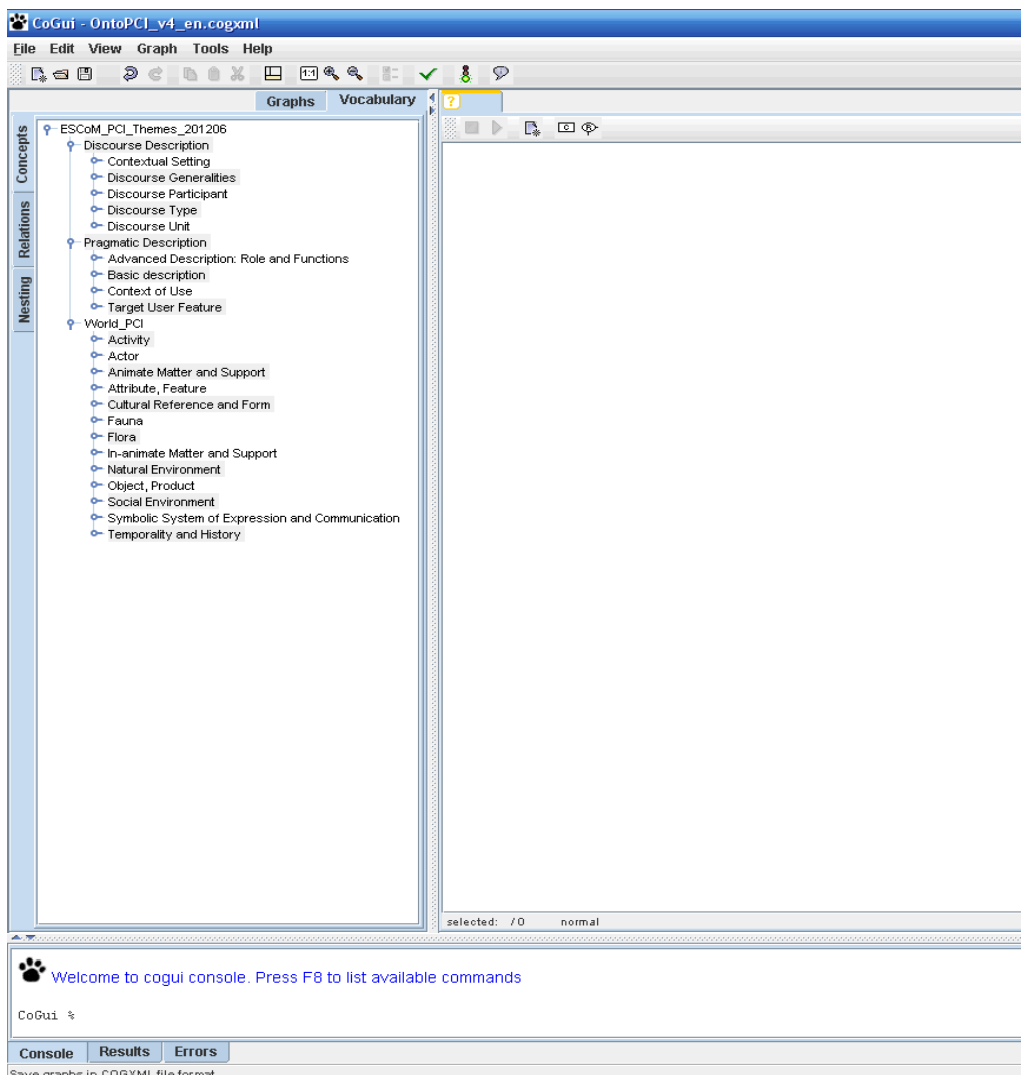

(*figure 3*: the overall picture of the PCI ontology of themes or concepts)

---

[16] cf. the web site of the CoGui project: http://www.lirmm.fr/cogui/spip.php?rubrique6
[17] Cf; the RCR web site: http://www.lirmm.fr/~mugnier/RCR/

The root ("ESCoM_PCI_201206") identifies the proprietary of the ontology (ESCoM), its abbreviated name and the date of its latest update (i.e.: the 20th of December 2006). The vocabulary of the PCI ontology is composed, as we can see it in figure 3, by three basic types:

1. the *concepts* or *themes* used for the description of the audiovisual records;
2. the *conceptual* or *thematic relations* (also called "*relational themes*") used for eliciting the relationships between concepts or themes developed in the audiovisual records,
3. the *nesting* or *contextual setting* of a specific type of thematic configurations (i.e. a set of themes linked between them).

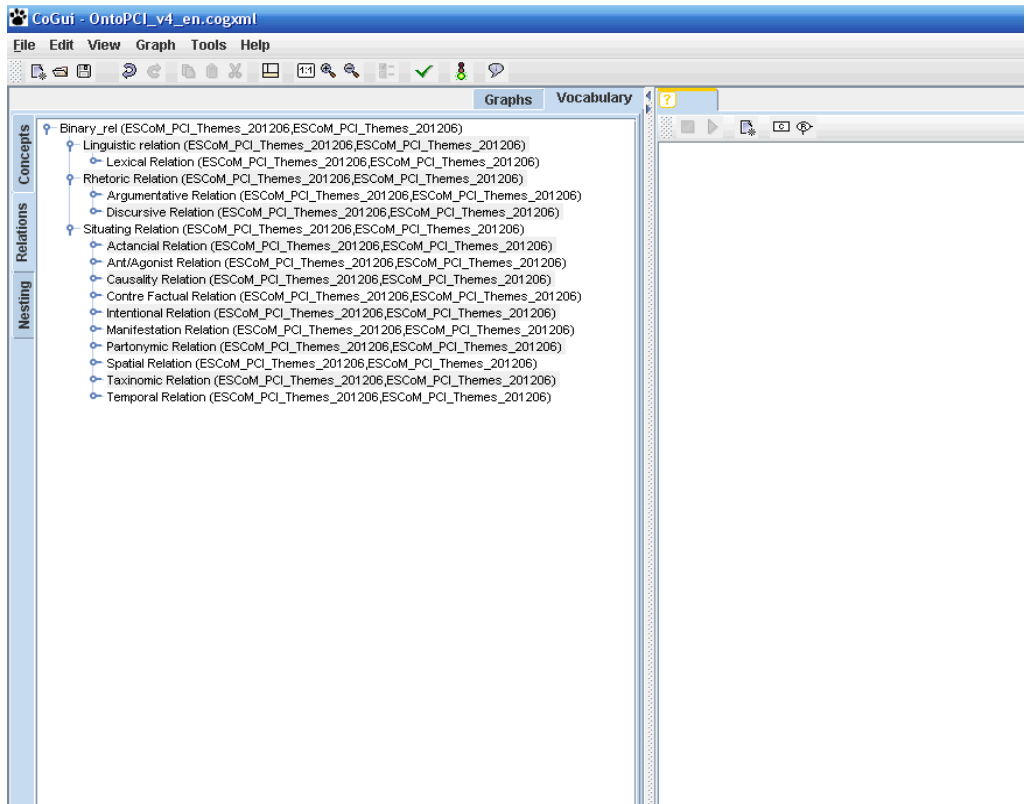

(*figure 4*: the overall picture of the PCI ontology of thematic or conceptual relations)

The whole vocabulary of the actually identified themes (concepts) is divided in three main upper categories:
1. the *referential domain* called the *"World_PCI"*;
2. the *narrative domain* of discourse production/interpretation called *"Discourse Description"*;
3. the *pragmatic domain* of the attuning of the uttered information called *"Pragmatic Description"*.

Remember that these three main upper categories are motivated with respect to the use of the PCI ontology, i.e.:
1. the *discourse analysis* of information uttered by an uttering subject (the researcher, the teacher, …) : the narrative or discourse domain;
2. with respect to a given referential domain, also called *knowledge domain* (in out case: cultural heritage) : the referential "world" domain;
3. where as the uttered information should be characterised with respect to *contexts and uses* for which it is especially relevant, for which it provides a maximum of added value: the pragmatic domain

These three upper categories will be presented and discussed in chapter 4.

A second part (figure 4) of the vocabulary of the PCI ontology is composed by the vocabulary of relations ("*properties*", in the CIDOC's CRM terminology[18]), the so-called conceptual or thematic relations or again relational themes. Main types of these relations are:

1. *Situating relations*, i.e. relations by the means of themes representing pieces of knowledge of especially (but not exclusively !) the referential domain (in our case: the World of PCI) are connected in forming *topical configurations* (corresponding to the "semantic graphs" in the above mentioned CWI projects[19]);
2. *Narrative relations*, i.e. relations by the means of which themes representing discourse knowledge (i.e. the production and organisation of – parts of – scientific discourse) are connected in forming narrative configurations;
3. *Linguistic relations*, i.e. relations by the means of which themes representing more specifically linguistic knowledge are interconnected in forming linguistic configurations.

The actual version of the three main categories of relations will be tested during the next months with respect to their interest of representing uttered knowledge of the PCI world. An improved version will be delivered at the end of 2007. A more detailed presentation of these three main types of relations will be given in chapter 6.

The third basic type composing the vocabulary of the PCI ontology refers to the nesting contexts, i.e. the grouping together of a configuration of themes (i.e. a "conceptual graph") which form the referential scope of one theme. In the actual version of the PCI ontology (version 4), only one nesting context has been defined, i.e. the *Discourse Topic* (figure 5). The *Discourse Topic* is the obligatory reference of any theme belonging to the *Discourse Type* hierarchy in the *Discourse Description* facet. In other words and simply spoken, a discourse selects and develops information of some referential domain (in our case: the PCI domain) – this information is called the topic of the discourse or again the discourse topic.

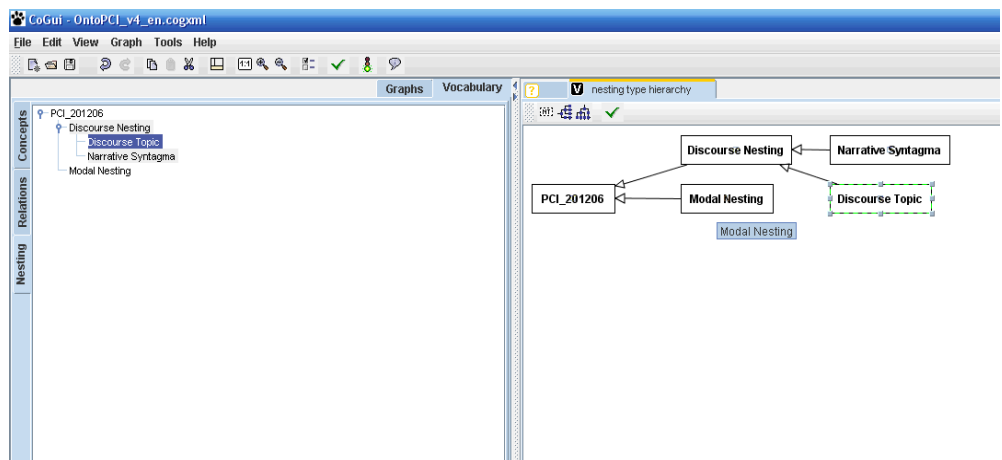

(*figure 5*: the overall picture of the nesting contexts specified in the PCI ontology)

Other nesting contexts will be introduced in the forthcoming versions of the PCI ontology, especially *modal contexts* (for representing beliefs, probabilities, obligations, etc.), *preferential contexts* (for representing appreciations and value judgements) or again *action contexts* (for representing the steps of performing an action).

---

[18] Cf. the online version of CIDOC's CRM, version 4: http://cidoc.ics.forth.gr/docs/cidoc_crm_version_4.0.doc
[19] Cf. footnotes 10 and 11

## 2.2/ The general structure of the conceptual graphs

With the help of the PCI vocabulary, conceptual graphs ("configurations"[20], in a more semantic or semiotic like terminology) can be specified. The objective of such conceptual graphs is to figure out knowledge patterns with respect to a given referential domain and in taking into account the fact that these knowledge is not an (in a classical sense) "objective" one but produced by an author, a speaker and that these same knowledge, at the same time represent a more or less relevant, important value or good for some destine.

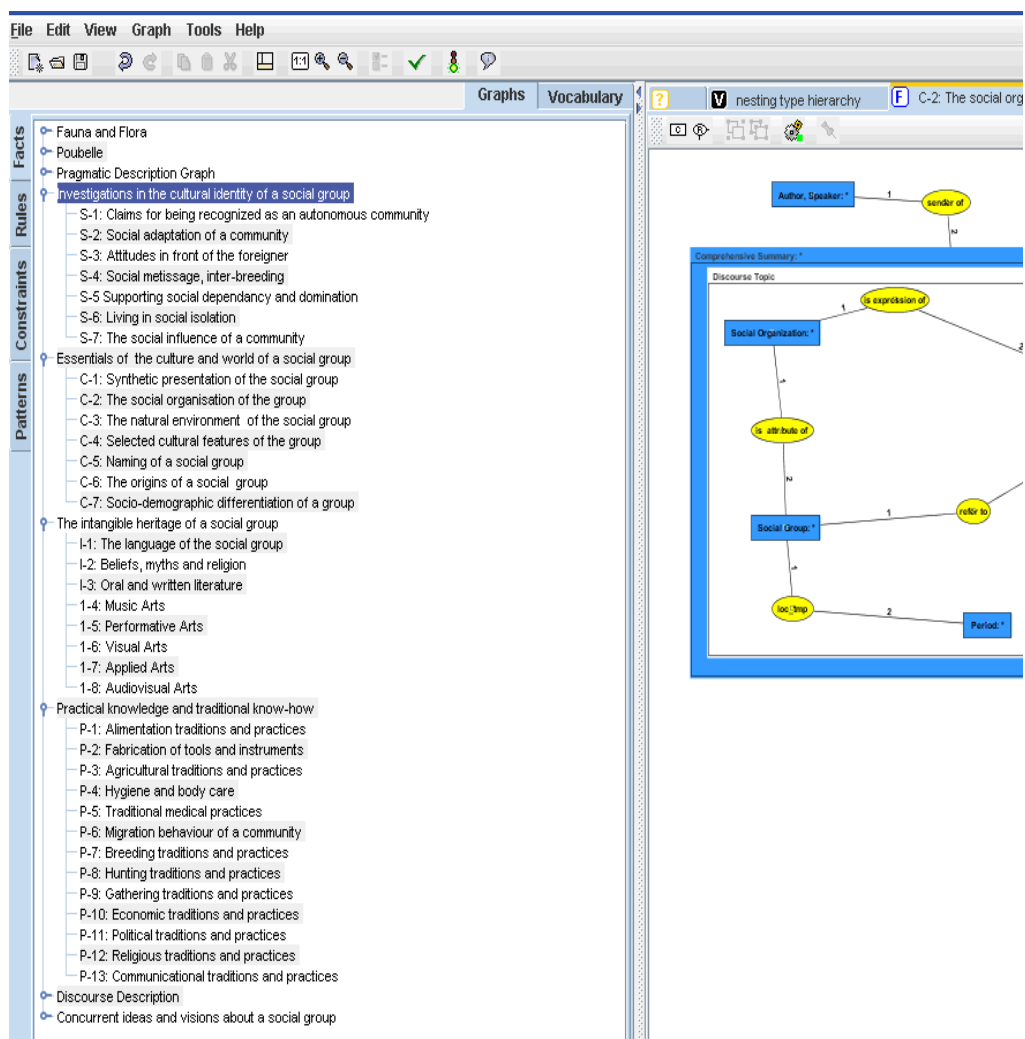

(*figure 6*: the overall picture of the PCI conceptual graph templates)

Concerning the PCI portal, this means, that conceptual graphs represent pieces of knowledge about the cultural heritage of specific social groups such as minorities or aborigines – knowledge produced and uttered by researchers, scholars, professionals or other people working in this domain which is or maybe could be of relevance for different user groups and specific contexts of use such as, for example, in education, research, cultural policy making, scientific journalism, etc.

---

[20] Cf. our working paper on *Conceptual analysis and knowledge management*, published in 1993 in Aarhus/Aalborg (Denmark) and Santiago (Chile), that develops the relationships between the structural analysis of discourse and text and conceptual graph theory: http://semioweb.msh-paris.fr/escom/ressources_enligne/p_stockinger/1992/extraction_textuelle.pdf

One specific use of conceptual graphs is to propose on the one hand *templates* for the indexing of audiovisual segments and on the other hand for identifying and exploring specific contents or topics in the field of cultural heritage of minorities and indigenous people. Figure 6 shows the actually elaborated catalogue of conceptual graphs which are used for the description and indexing of audiovisual segments belonging to the PCI corpus.

## 2.3/ Terminological questions

*Themes* are expressed in English language by *nouns* or *noun phrases* (*nominal groups*) using *title case* (*initial capitals*). As far as possible, nouns or noun phrases are in *singular form*. *Relations* are generally expressed by *verbal phrases* in *present tense*. Finally, *nesting contexts* are expressed - like themes - by *noun phrases* in *singular form*.

Let us mention that the actual version of the PCI ontology (version 4) doesn't possess a single enumeration system for themes, relations and nesting contexts. Such a system will be provided for the next version which will identify without ambiguity:
- Themes by a number preceded by the capital T (example: T-1: Discourse Description)
- Relations by a number preceded by the capital R (example: R-1: Linguistic Relation)
- Nesting Contexts by a number preceded by the capital C (example: C-1: Discourse Nesting).

It has to be emphasized that we employ a terminology more closer related to text and language studies than to (predicate, …) logics (the reference of conceptual graph theory). In this sense, we use:

- the term "theme " for the term "concept corresponds to the term "theme",
- the term "thematic relation" for the term "conceptual relation",
- the term "referent" or "key word" for the term "instance" or "value" (of a concept)
- and the term "thematic configuration" for the term "conceptual graph".

The terms "theme" and "concept" corresponds to the term "class" (or "entity, in previous versions) in CIDOC's CRM (version 4)[21] and the term "thematic, conceptual) relation" to the term "property".

Finally, it also has to be emphasized some definitional questions concerning themes, relation between themes, referents of themes as well as thematic configurations.

Themes are *generic concepts*. For instance the theme [Minority:*] circumscribes a certain type or category of social groups that share features following which they can be considered as a minority. The referent is a *value* or an *instance* of a theme. For instance, the instanciated theme [Minority: walachians] means that there is a social minority called walachians.

For a very comprehensive discussion of these issues, we refer to the *Categories of the Description of Works of Art*, edited by Murtha Baca and Patricia Harpring[22] on the behalf of the J. Paul Getty Trust. For technical explanations, we strongly recommend Sowa's reference work on *Conceptual Structures*[23].

It has to be stressed that the referents of the PCI ontology are "*words*" uttered by a speaker, i.e. his/her *saying about something*, in our case, his/her saying about something in the domain of cultural heritage of minorities and indigenous people. This means that the referent of the PCI ontology are –

---

[21] Cf. the online version of CIDOC's CRM, version 4:
http://cidoc.ics.forth.gr/docs/cidoc_crm_version_4.0.doc
[22] Cf. the online documentation:
http://www.getty.edu/research/conducting_research/standards/cdwa/index.html (and especially the chapter: "Generic Concept Authority")
[23] Cf. J. Sowa, *Conceptual Structures: Information Processing in Mind and Machine*. New York, Addison & Wesley 1984

broadly speaking – *linguistic* (more precisely "semio-linguistic", i.e. *uttered verbal*, *shown visual*, ...) entities. The referent is not – at least in the sense as we use it here – some individual or group of individuals "out there".

To stress this point doesn't mean only to circumscribe the field of use of the PCI ontology (i.e. the description of the **content of discourses** dedicated to a specific domain of knowledge) but it makes clear also the status of *key words* and the description of discourses by *key words*.

*Key words* are the **values** or **instances** of a theme. For instance "walachians" is a key word of the generic theme [Minority:*]. As well known, there are two central categories of key words:

1. *free key words*;
2. and *controlled vocabularies*.

Free key words are produced by the indexer "freely" (i.e. without any reference to a controlled vocabulary) but following the structure of the ontology (especially the vocabulary composing the themes of the ontology) and in respecting some formal instructions defined, in the case of the PCI pilot, by an internal technical document used for the indexing of a video belonging to the AAR Programme[24]. Free key words can be – and, in the case of the PCI pilot, are – produced in different languages: English, French, Spanish, Russian, Polish, Italian, etc.

Controlled vocabularies are lists of terms ("key words") which have been approved by a community (an institution, ...) and which are used such as. Controlled vocabularies compose the body of thesaurus (hence the close association between thesaurus building and use or reuse of controlled vocabularies[25]) but controlled vocabularies also can be glossaries, terminologies or specialised lexica. For the PCI purpose of importance is, for instance, the ISO 639.2 list of natural language names[26]. Another valuable source is the Wikipedia List of people[27], i.e. individuals (by name, by nationality, etc.). An important source for the PCI pilot is the list of indigenous people produced by the *NGO The Peoples of the World Foundation*[28].

Both – free key words and controlled vocabularies – are used in the PCI pilot. Nevertheless, the free key wording constitutes the more important description and indexing activity. It is done *manually* and recovers two principal activities:

1. the *term extraction activity* (i.e. the extraction of a term or a series of terms that constitute the referent of a theme and a part of the discourse topic);
2. the *content paraphrasing* (i.e. the production of a small paraphrase condensing the referential content of a theme and forming a part of the discourse topic).

There are techniques – especially in the field of "speech to text transcription", text mining and knowledge management – that seem to be able to reduce the manual work of indexing but they cannot be used as such in this project[29].

---

[24] http://semioweb.msh-paris.fr/AAR/FR/
[25] Cf. the web site dedicated to controlled vocabularies and thesaurus building of the Queensland University of Technology: http://sky.fit.qut.edu.au/~middletm/cont_voc.html
[26] Cf. http://www.loc.gov/standards/iso639-2/php/French_list.php
[27] Cf. http://en.wikipedia.org/wiki/Lists_of_people
[28] Cf. the web site of the Peoples of the World Foundation: http://peoplesoftheworld.org/ and especially the interactive list of indigenous people in the world: http://peoplesoftheworld.org/bypeople
[29] For this, cf. the Sail Labs Technology homepage: http://www.sail-technology.com/

# 3/ The building of the PCI ontology

## *3.1/ Methodology and guiding principles*

As already mentioned, the PCI ontology is conceived especially for the description and indexing of *uttered knowledge*, i.e. of knowledge produced in a discourse by someone for someone else … This means that it has:

- to circumscribe and positively elicit the domain of reference, i.e. the *semantic universe* of the produced discourse (also called the "universe of discourse");
- to elicit the different parts and components composing the discourse;
- to elicit also the (possible) target of a discourse, i.e. its destine, its context of use and here mainly its publishing environment.

Conception, specification and building the PCI ontology have followed an empirical approach that relies heavily on the *specificity of the PCI source corpus* itself, i.e. on the *produced discourses* by researchers and other specialists. The *ten* main steps of the building and maintaining of the PCI ontology are:

> 0/ Definition of objectives and constitution of a pilot committee.
> 1/ Empirical work on the corpus.
> 2/ Familiarization and use of an ontology editor.
> 3/ Work on pre-existing literature and initiatives.
> 4/ Building of first small versions and local testing.
> 5/ Building of a first more stable general version (previous version 3 and actual version 4).
> 6/ Working with the first stable version on the source corpus.
> 7/ Note taking of limits, lacks, degree of usefulness, degree of comprehensibility of the first stable version.
> 8/ Updating of the first stable version (version 5).
> 9/ Generalizing of the PCI ontology (its partial use in other domains, its comparability with respect to other ontologies, issues of interoperability, etc.).

**Step 0/ Definition of objectives and constitution of a pilot committee**

The general objectives of the PCI ontology have been specified in a previous ESCoM report[30] concerning the authoring and learning scenarios that will have to be realised within the Logos project. There are five central scenarios concerning the re-authoring or again the repurposing of existing audiovisual digital content in taking into account especially the central mission of the French Audiovisual Archive Programme in Social Sciences and Humanities (AAR)[31], viz. the opening of the existing digital content to a multilingual and multicultural community. The pilot has been defined in another report[32] as belonging to the sector of cultural heritage of indigenous people and minorities.

---

[30] Peter Stockinger, Alice Maestre, Elisabeth de Pablo and Anaïs Debaisieux: *The four ESCoM-FMSH learning scenarios for the IST project LOGOS* , Paris 2006: http://semioweb.msh-paris.fr/escom/ressources_enligne/projets_recherche/06_09_logos/WP2_2_Scenarios_Logos_ESCoM.pdf
[31] Cf. the portal of the AAR Programme: http://semioweb.msh-paris.fr/AAR/FR/
[32] Peter Stockinger, Alice Maestre, Elisabeth de Pablo and Vincent Dupont: *Audiovisual Archive materials for the Logos project from ESCoM (Equipe Sémiotique Cognitive et Nouveaux Médias) in Paris*, Paris

A pilot committee has been constituted of which the principal missions are the realisation and maintenance of the PCI web portal, the communication of the web portal to the interested communities (especially in education, research, journalism and NGO), the participation in the constitution of the relevant source corpus and its evolution (i.e. the planning and realisation of new events to be filmed, events such as interviews, exhibitions, research seminars, documentaries, etc.), the participation in the elaboration and maintenance of the PCI ontology understood as a means for the description of the uttered knowledge in the field of cultural heritage of minorities and indigenous people, the participation in the description and indexing work of the audiovisual corpus, the contribution in the realisation and testing of the publishing and learning scenarios, and, finally, the participation in the promotion and dissemination of the concrete results of the PCI pilot.

The pilot committee comprises actually *8 persons* composed by researchers, educationalists and professionals working in the fields of anthropology, ethnology and sociology, information and communication sciences, archive and librarian sciences, semiotics and text analysis as well as either in universities or high schools.

### Step 1/ Empirical work on the corpus

As already mentioned, the source corpus of the PCI portal is composed of a selection of about 100 hours videos belonging to the 2500 hours audiovisual resources of the AAR Programme[33]. The principal activities in this first step which is yet ongoing, are:

1. The repeated visioning of the PCI source corpus.

2. The rough description and analysis of this source corpus by the means of usual semiotic or text-linguistic tools (identification and enumeration of principal topics, identification of main discourse genres used for uttering the topics, general characterisation of the discourse style, evaluation of the degree of specificity of the content proposed, …).

3. A specific emphasis of this rough analysis has been given to the reuse of parts of the produced discourses, and this again by usual semiotic or pragma-linguistic tools (trying to satisfy typical questions such as the produced discourse could be reworked/repurposed *for whom?* for *which context?* by the means of which *publishing genre?*)

Technical means used for this work on the corpus is a audiovisual segmentation and semiotic description tool co-developed by INA Recherche and FMSH-ESCoM. The name of this tool is *Interview*[34] and it is used, within the context of the AAR Programme, for the publishing of multilingual versions of an interview or conference[35].

---

2006: http://semioweb.msh-paris.fr/escom/ressources_enligne/projets_recherche/06_09_logos/WP2_1_Archive_material_ESCoM.pdf

[33] Peter Stockinger, Alice Maestre, Elisabeth de Pablo and Vincent Dupont: *Audiovisual Archive materials for the Logos project from ESCoM (Equipe Sémiotique Cognitive et Nouveaux Médias) in Paris*, Paris 2006: http://semioweb.msh-paris.fr/escom/ressources_enligne/projets_recherche/06_09_logos/WP2_1_Archive_material_ESCoM.pdf

[34] For a short presentation of the Interview tool and its interest for the repurposing of digital content, cf. Peter Stockinger, Semiotic Video Processing and Personalised Publishing, Paris 2005: http://semioweb.msh-paris.fr/escom/ressources_enligne/projets_recherche/04_08_ina/conference_10_05.pdf

[35] Cf. the online document explaining the Interview tool: http://semioweb.msh-paris.fr/escom/ressources_enligne/projets_recherche/04_08_ina/conference_10_05.pdf
cf. also the use of this tool in teaching activities dedicated to the production ("versioning") of multilingual versions of audiovisual files: http://semioweb.msh-paris.fr/escom/fr/enseignement/annees/06_07/lisbonne_07.htm

The result of this systematic empirical investigations on the PCI corpus is a first structured lists of *lexical items* representing the topics as well as associated to this items, *rhetorical annotations* (discourse genre; etc.).

It has to be stressed that this list of lexical items constitute not only a first input for the building of the PCI ontology but also its principal input of it given the specific requirement for the PCI ontology to be priory a description and indexing tool of the produced knowledge in this corpus.

### Step 2/ Technological environment

In order to transform progressively the (evaluated, refined, …) list of lexical items obtained a s a result in step 1, an ontology editor has been chosen which is the CoGui ontology editor belonging to the Cogitant environment for working with conceptual graphs.

This ontology editor fits well with the some more specific theoretical assumptions in *structural* semantics and discourse analysis stressing more particularly the fact that a list of items are not enough to produce a coherent and interesting description of the meaning of a discourse or any semantic field. For this, schemas or lexical (thematic, rhetoric, …) configurations have to be defined representing the specific position of a lexical item (a theme, …) with respect to other lexical items (themes) used in a description. This assumption can be easily "translated" in the terms of "conceptual graphs"[36].

CoGui requires, more particularly, a specific form of building an ontology. Indeed, an ontology is composed by:

1. a vocabulary of *concepts*,
2. a vocabulary of *relations*(restricted or not in their scope of applicability),
3. a vocabulary of *nesting contexts*,
4. and a structured list of *conceptual graphs*, *rules*, etc.

Given this formal environment, the activities of conceiving and building an ontology - such as the PCI ontology – is constrained and oriented. But it has to be emphasized once more, that the restrictions imposed by the CoGui ontology editor are in very harmony with the structural approach of discourse analysis in linguistic and especially in semiotics – contrarily to many other ontology editors …

### Step 3/ Pre-existing literature and initiatives

In order to assess the list of lexical items obtained in step 1, a specific task has been to investigate already existing initiatives, comparable with the scope and the objective of the PCI ontology. This means that a certain number of specialised literature and existing terminological resources (ontologies, thesaurus, …) has been examined and - as far as possible - re-used for the building of the PCI ontology. Specialised literature and already existing resources have been taken into account especially in (immaterial and material) cultural heritage, in social sciences related to studies of people and communities of people as well as in discourse analysis and description. These investigations will be presented here after (cf. 3.2: ontologies, thesaurus and terminologies).

There are three main issues in studying critically pre-existing literature and other already existing resources such as controlled vocabularies:

1/ to proceed to the identification of the principal upper categories of the ontology (the *taxemes* in a semiotic oriented discourse theory and terminology), i.e. to answer the always problematic question of "what are the ground categories of my ontology?";

---

[36] cf. note 18

2/ to understand the principal lacks in the already established list of lexical items (i.e. which constitutes one of the main result of the activities of step 1) and to evaluate if these lacks are really relevant for the scope and purpose of the ontology;

3/ to prepare also the maintenance of the ontology, i.e. the "monotonic" completion of a given vocabulary by a new vocabulary (which either introduces a new concept type hierarchy or expanses an already given concept type hierarchy).

A particular problem here is the question of the interoperability with other related ontologies. The interoperability is *not only a formal problem* but much more a problem from the point of the view of the *adopted theory* by the means of which the – especially – basic structure of a vocabulary is elaborated (cf. the question of the identification of the upper categories or taxemes).

For instance, CIDOC CRM version 4 is very different from this point of view of the PCI ontology whereas certain thesauri adopt a rather similar point of view. For instance Getty's AAT and, naturally, UNESCO's thesaurus in social and human sciences adopt upper categories which are highly congruent with the upper categories of the referential facet "World_PCI".

GOLD[37], a well known ontology for language and discourse studies, possess a structure which directly fits with one of the themes in the *Symbolic System of Expression and Communication hierarchy* of the PCI ontology: the theme "*Linguistic Structure*" in the PCI ontology is synonymous with one of GOLD's major upper categories – the *Linguistic feature* category. A "monotonic" evolution of a part of the PCI ontology incorporating the possibility to describe linguistic topics, is therefore possible but, for the moment at least, not relevant.

### Step 4/ Building small versions or pieces of the intended ontology

The building of an ontology (like any other description tool or means) is necessarily based on some implicit hypotheses. In our case, there are the two following hidden hypotheses:

1. hypothesis on the narrative and rhetorical structure of language and discourse as the principal knowledge production, communication and appropriation means (concerns mainly the *Discourse Description* facet of the PCI ontology);
2. hypothesis on the *stereotypical organisation* of the referential knowledge domain – hypothesis related to the phenomenological notion of "*lifeworld*"[38] (E. Husserl, A. Schütz) and the central notion of action or activity and implied participants localisable in space and time and making sense with respect to a conceptual framework called *culture*.

Given such "assumptions" and with respect of the results of previous lectures of the audiovisual corpus as well as with respect to already existing literature and other, comparable initiatives, small pieces of – especially thematic and relational – vocabulary of the PCI ontology have been elaborated, i.e. especially:

- the *Actor* theme hierarchy;
- the *Activity/Social Practice* theme hierarchy;
- and a subset of well-known types of relations called *actantial* or *casual relations* allowing the identification of a role of a participant in an activity or social practice.

These themes and relations represent a very partial coverage of the domain of cultural heritage, i.e. all knowledge concerning mentionable activities in social communities presented and discussed by the contributors to the PCI source corpus.

---

[37] cf; the GOLD community web site: http://www.linguistics-ontology.org/gold.html
[38] for a technical use of lifeworld analysis in computer simulation programmes, cf. for instance Philip Agre and Ian Horswill, Life World Analysis ; San Diego/LA Jolla & Evanston 1997:
http://www.cs.cmu.edu/afs/cs/project/jair/pub/volume6/agre97a-html/lifeworlds.html

Small graphs have been elaborated, based on the actantial (casual) positioning between themes belonging to the Actor hierarchy and themes belonging to the Activity hierarchy. These graphs are description models of social practices implying the participation of members of a social actor. They have been tested with respect to the audiovisual material.

In a second time, investigations for broader empirical coverage of the PCI ontology have been undertaken. Concretely speaking, other hierarchies of themes have been elaborated such as, for instance:
- the *Attributes* and *Features* theme hierarchy;
- the *Object* and *Product* theme hierarchy,
- the *Temporality* and *History* theme hierarchy,
- and especially the *Discourse Description* theme hierarchy as well as the *Pragmatic Description* theme hierarchy.

### Step 5/ Building of a first more stable general version (version 4)

Progressively, out of the different "local" pieces (Step 4), a first more stable version of the PCI ontology has been created which is the result of the fusion of the different local pieces elaborated and tested in the previous step.

The first stable version has been version 3.5, released and sent to the Logos consortium at the end of November 2006. This report is based on an updated version of the PCI ontology - version 4. This means that the claim of these last two versions of the PCI ontology are to provide:

1. a vocabulary that covers at least on the top (i.e. not always in a more fine grained way) very large parts of the referential domain – the domain of cultural heritage of minorities and indigenous people;
2. a vocabulary that allows a rough description of the discourse activities by the means of which knowledge of the referential domain is produced and transmitted to a destine;
3. a vocabulary that allows at least a rough description of the (potential) reuses of the uttered knowledge with respect to a given user profile and a given context of use.

This means: no claim is yet made for a whole coverage. For example, the two yet very general themes "*Social organisation*" and "*Social field*" (belonging to the *Social Environment* theme hierarchy) are not elaborated, only indicated (like, for instance, the categories – or entities – "*time span*", "*dimension*" or "*place*" in the 4th version of the CRM of the CIDOC). This is also true for the two theme hierarchies "*Fauna*" and "*Flora*" of which the deployment will be realised progressively only in accordance with the specific descriptive of our corpus and in accordance to controlled specialised (botanical and zoological) vocabularies or thesauri.

This is also true concerning the actual state of the relational theme hierarchy: the upper categories are supposed to be stable but details have to be decided again as well as the more restrictive definition of the scope of each relation (i.e. its applicability with respect to types of themes).

### Step 6/ Working with a stable general version

With the PCI ontology, version 4, the audiovisual source corpus will be reworked during a large period of 2007 – a reworking based on the previous empirical investigation (Step 1) and which is concerned more particularly by:

- the more fine grained (re-)segmenting of the corpus;
- the more detailed description and translation of the audiovisual segments;
- the indexing of each audiovisual segment with the help of the ontology of referential themes (selection of one or more themes in the ontology and, eventually, key wording of a selected theme where "key word" = instance of the theme);
- the indexing of each segment by the means of the discourse description theme hierarchy;

- the annotation of each segment by means of the pragmatic description theme hierarchy (which will be replaced latter on by the pragmatic description tool of the Logos authoring Studio);
- the indexing of each segment with the help of indexing templates (i.e. conceptual graphs).

The result should be a whole described and indexed audiovisual corpus ready for the republishing (repurposing) following the publishing and learning scenarios defined in the already above mentioned ESCoM report[39]. A first part of this result is projected to be available at the end of September 2007; a second part at the end of May 2008.

### Step 7/ Note tacking of limits of the actual version of the PCI ontology

The description and indexing process of the audiovisual corpus by the means and with the help of the actual (4th) version of the PCI ontology will more or less probably uncover:

- *lacks* in the vocabulary for adequately describing topics or strategies for uttering topics or again for adequately describing the repurposing of an uttered topic;
- the degree of *usefulness* of given hierarchies of themes for concrete descriptions (themes which are too specialised, themes which are, contrarily, to general, themes that give too partial vision of an intended referential object, etc.);
- the degree of *comprehensibility* of the meta-linguistic expressions of themes.

This information are gathered in a sort of a *quality evaluation document* of the actual version of the PCI ontology and will serve for the next updating of the ontology in the last trimester of 2007.

Meanwhile parallel investigations in the structure of the actual version of the ontology will be undertaken by the means of technical meetings with experts and the pilot committee as well as the critical and comparative analysis of already existing resources.

### Step 8/ Updating of the actual version

In the last trimester of 2007, a new updated version of the PCI ontology will be released. It will be the result of a concrete work with the actual version 4 on a material of almost 60 hours of video and a critical evaluation of its descriptive limits and its usefulness.

### Step 9/ Generalizing of the ontology

It is expected that the next version (version 5) of the PCI ontology will constitute already a very stable version of which specific parts could easily be reused. For instance, the facets "*Discourse description*" and "*Pragmatic Description*" should be become reusable in any enterprise aiming at the description of *uttered knowledge*.

Also parts of the referential facet should become reusable in other descriptive enterprises. This is certainly the case for the *Activity theme hierarchy* as well as for the *Culture theme hierarchy* or again the *Actor theme hierarchy*.

For this, new pilots will be identified and developed – already during the year 2007. One pilot will be developed in cooperation with a French Publisher on the "Making of Europe" based on audiovisual records of international symposia in Bordeaux, Paris, Prague, Roma, Madrid and Istanbul dedicated to European Community. Another pilot will be developed with the University of La Sapienza in Roma on critical media studies based on audiovisual records of seminars, workshops and interviews dedicated to the media industry. Other pilots, closer to the PCI pilot, are actually under negotiation.

---

[39] Peter Stockinger, Alice Maestre, Elisabeth de Pablo and Anaïs Debaisieux: *The four ESCoM-FMSH learning scenarios for the IST project LOGOS*, Paris 2006: http://semioweb.msh-paris.fr/escom/ressources_enligne/projets_recherche/06_09_logos/WP2_2_Scenarios_Logos_ESCoM.pdf

## 3.2/ Already existing resources: thesaurus, terminologies and ontologies

In having in mind the specific orientation of the PCI ontology (i.e. to be a description/indexing tool of *discourse topics* related to cultural heritage issues of minorities and indigenous people, it is clear that its elaboration and maintenance are undertaken in a comparative way, i.e. in referring to already existing more or less closely related initiatives.

In our case, this means especially the evaluation of already existing ontologies, thesauri, terminologies and other types of vocabularies that at least partially have to do with the object of the PCI ontology. In this case, already existing resources have been looked for and critically analysed that are related either to the complex "*World_PCI*" or to the complex "*Discourse Description*". The main objectives of these comparative studies have been and are the following ones:

1. understanding of the basic categorisation of the two (referential and narrative) facets of the PCI ontology (the Discourse Description and the PCI World as well as of the third (pragmatic) facet, the Pragmatic Description) with respect to already existing conceptual decisions;
2. reuse of already existing vocabularies;
3. interoperability with already existing initiatives.

Naturally, these three objectives are determined by the main objective, *viz.* the production of a description/indexing tool for a (open) audiovisual corpus composing the PCI web portal. In other words, the determining aspects for the categorisation and composition of the PCI vocabulary are:

1. the empirical specificity of the content of the audiovisual corpus;
2. main assumptions and theories in semiotic and linguistic discourse and text description and analysis.

For the complex "*Discourse Description*", several resources have been identified and studied in order to find an organisation form of the narrative facet of the PCI ontology which on the one hand fits with the special requirements concerning the PCI ontology (i.e. to be a description and indexing tool for discourse topics), and which on the other hand is sufficiently open and general for being able to integrate already existing propositions and "standards":

- the *Text Encoding Initiative* (TEI)[40] especially with respect to the analysis of specific discourse features: setting, discourse types, etc;

- the *General Ontology for Linguistic Description* (GOLD)[41] and especially its central category of the "Linguistic Features", which is called in the PCI ontology *Linguistic Structure* and which can be "monotonously" developed within the PCI ontology if there is some need for describing more detailed language features, linguistic documentation, etc.;

- Narrative ontologies[42] especially with respect to the development of the *Discourse Unit* theme specialising a branch in the *Discourse Description* theme hierarchy (narrative facet of the PCI ontology) as well as the *Narrative Relation* taxeme belonging to the relational part of the PCI ontology

---

[40] Cf http://www.tei-c.org/
[41] Cf. the GOLD community web site : http://www.linguistics-ontology.org/gold.html
[42] Cf. the very interesting on narrative ontologies of Henrik Scharfe of the Aalborg University (Denmark): http://www.hum.aau.dk/~scharfe/cv.htm. It has to be noticed that H. Scharfe works with another conceptual graph ontology builder (the Amine ontology builder, developed by Adil Kabbaj of the INSEA in Morocco): http://www.huminf.aau.dk/cg/index.html

Concerning the referential facet "World PCI", many resources have been consulted in order to structure better specific aspects of the World PCI referential taxeme. But almost all existing resources are - at least concerning the particular objective – sometimes very different from that of the PCI ontology, *viz.* to be a discourse topic description and indexing "language".

For instance, the CIDOC's CRM[43] is supposed to be an indexing tool for museum information systems, no for discourse analysis. Also, it interprets "cultural heritage" in a very restrictive sense and not in the sense used for the PCI pilot, *viz.* in the sense of social sciences as (collective) knowledge and value systems and processes of social actors inhabiting a social and historical as well as natural world. Moreover the CIDOC CRM uses upper categories for their *classes* (i.e. concepts or themes) that are motivated by a sort of a temporal philosophy, a *temporal* like theoretical framework. There are, as it seems, five basic or upper categories from which actually only two are really developed: the *Time entity* and the *Persistent Item entity*. Both represent deep hierarchies which are not always very easy to understand (from a purely notional point of view) but they show the "underlying philosophy" which refers to some sort of temporal common sense reasoning: there are things that change in time and things that don't change…

In any case, for dealing with the referential facet *World_PCI*, this reference to a sort of temporal philosophy as it prevails in the CIDOC CRM is a far too abstract approach which doesn't share any relationship with the traditional approach of "life worlds" or cultures in social and human sciences.

A common reasoning based on a sort of "life world" conception is, contrarily and surprisingly, often present in *thesaurus like resources*, such as the UNESCO thesaurus in social and human sciences[44], the AAT[45], the IconClass[46] thesaurus, the European Heritage Network (H.E.R.E.I.N.)[47] project and its multilingual thesaurus based on 9 "life world like" basic concepts for dealing more particularly with architectural and archaeological heritage, etc..

In comparing these and other thesauri, it is evident that there doesn't exist any explicit agreement concerning the basic categories – the taxemes or again the upper categories - that could or should, so to speak, constitute a *canonical basis* for building thematic hierarchies dedicated to the description of cultural heritage (in a restrictive museum like sense as well as in a more broader, social sciences sense). But there is a sort of implicit, tacit convergence between different thesauri in opting for such taxemes or basic or again upper categories. A very good example is the AAT built on taxemes or basic categories such as:

- Agents,
- Physical attributes,
- Activities,
- Materials,
- Objects,
- Abstract concepts
- Styles, Periods, and Cultures
- etc.

With some imagination, it is rather easy to find correspondences between these upper categories or taxemes and those that organise the referential basis of the World_PCI. This is simply a proof that both are sharing a sort of implicit reference model sometimes called in a phenomenological tradition "life world" (cf. Husserl, Schütz). Also the CIDOC the "highest" categories of the CIDOC represent more or less the same picture:

---

[43] Cf; http://cidoc.ics.forth.gr/
[44] Cf; http://www2.ulcc.ac.uk/unesco/4.htm
[45] Cf. the Getty's web site: http://www.getty.edu/research/conducting_research/vocabularies/aat/
[46] Cf; the IconClass Libertas homepage: http://www.iconclass.nl/
[47] Cf the HEREIN web site: http://www.european-heritage.net/sdx/herein/

- Places,
- Events,
- People,
- Things,
- Concepts[48].

But it is strange, that these – intuitively clear – categories disappear in the conceptual model of classes or entities that compose the CIDOC CRM.

Let us also mention also the 10 main categories of the IconClass Thesaurus[49], especially the categories 1 to 6 which are, once more again, very close related to the above mentioned ones as well as to the basic themes or taxemes composing the referential facet of the PCI ontology:

- Religion and Magic,
- Nature,
- Human Being,
- Society,
- Civilisation and Culture,
- Abstract Ideas and Concepts,
- History.

The selection of a canonical basis of taxemes for structuring and classifying the themes of the referential facet of the PCI ontology, called *World_PCI*, therefore refers strongly to the life world model as conceived and discussed especially in phenomenology as well as in a specific type of philosophy of language (the "language game" philosophy of L. Witgenstein adapted by scholars to researches in social and human sciences).

For the explicit deployment of each theme hierarchy, helps and hints have been used coming either from specialists and experts or from specialised literature or again from already existing thesauri and other forms of "controlled vocabularies" such as terminologies, glossaries, etc.

A very important reference is the UNESCO thesaurus of social and human sciences[50] composed by basic themes such as:

- "Population",
- "Family",
- "Ethnic question",
- "Social systems",
- "Human settlements and land uses,
- "Social policy and welfare",
- "Social problems".

Different of them have been introduced in the corresponding theme hierarchy of the referential facet of the PCI ontology. The specialised themes deploying the UNESCO's categories have been selectively reintroduced in the PCI ontology with respect to their usefulness for empirical description and indexing needs of the audiovisual corpus of the PCI pilot. UNESCO's thesaurus has been very useful for organising the *Actor* theme hierarchy, the *Activity* theme hierarchy as well as the *Temporality, History* theme hierarchy.

---

[48] Cf. CIDOC Relational Data Model:
http://www.willpowerinfo.myby.co.uk/cidoc/model/relational.model/datamodel.pdf; cf. also the EC project ArteFact: http://www.art-e-fact.org/ARTEFACT_may_2004/NEWRELEASE/index.php?concept

[49] Cf. the IconClass Libertas Browser: http://icontest.iconclass.nl/libertas/ic?style=index.xsl
[50] UNESCO Thesaurus of social and human sciences: http://www2.ulcc.ac.uk/unesco/4.htm

A highly valuable resource has been The British Museums Object Names[51] thesaurus of which the list of top terms has been used for the organisation of the basic referential theme *Object, Product*. This resource has been completed by another, very suggestive ethnographic thesaurus – the well known SVCN (*Stichting Volkenkundige Collectie Nederland*)[52] thesaurus (in Dutch). In order to organise a specific but very important branch of the Object, Product hierarchy – the Arts Work theme hierarchy - the *Boekmanstichting Studiecentrum voor kunst, cultuur en beleid,* a well known Dutch thesaurus of Cultural policy[53] has been consulted an extensively used.

For elaborating the *Fauna* and the *Flora* theme hierarchy, the central reference has been (and continues to be) the as impressive as beautiful *Tree of Life Web* project of the NSF and the University of Arizona[54]. The progressive elaboration of the *Inanimate Matter* theme hierarchy will strictly follow the empirical description and indexing needs of the corpus of the PCI pilot and be based on the one or the other of the generally admitted references in this field such as the Strunz classification of solids[55].

Finally, let us also mention two very valuable help tools for specifying progressively a thematic hierarchy – tools which have been used extensively for building the PCI ontology and which will be used in the coming months for the realisation of the new version 5:

- The visual thesaurus of ThinkMap[56].
- and, naturally, the well known and extremely useful WordNet Online[57] of the Princeton University.

---

[51] Cf. the British Museums Object Names Thesaurus: http://www.mda.org.uk/bmobj/Objintro.htm
[52] Cf. the online version of the SVCN thesaurus (in Dutch) : http://www.svcn.nl/thesaurus.asp
[53] Cf.: http://www.boekman.nl/download_thesaurus.html
[54] cf. the web site of the Tree of life web project: http://tolweb.org/tree/phylogeny.html
[55] Cf; http://webmineral.com/strunz.shtml
[56] Cf. http://www.visualthesaurus.com/trialover.jsp
[57] Cf. http://wordnet.princeton.edu/

# 4/ The thematic hierarchies of the PCI ontology

As already shown in chapter 2, the three taxemes (or upper categories) of the PCI ontology (more precisely: of the actual version 4 of this ontology) are:

1. the taxeme *World_PCI*;
2. the taxeme *Discourse Description*;
3. the taxeme *Pragmatic Description*.

These three taxemes or upper categories of the PCI ontology are motivated with respect to its specific use as a description tool:

1. of the *discourse of information* uttered by an "uttering" subject (the researcher, the teacher, …) : *the narrative domain*;
2. with respect to a given referential domain, also called *knowledge domain* (in out case: the cultural heritage) : the *referential "world" domain*;
3. whereas the uttered information should be characterised with respect to *contexts and uses* for which it is especially relevant and provides a maximum of added value: the *pragmatic domain*.

These three taxemes depend on the root category labelled *ESCoM* and containing temporal information concerning the latest update of the ontology.

## 4.1/ The narrative facet of the ontology: the description of a produced discourse

The taxeme *Discourse description* (figure 7) is specialised in five basic themes that cover, hypothetically, the principal aspects and features of the production of a discourse (in our case: of a scientific discourse, *lato sensu*):

1. **Contextual setting**. Theme that is itself deployed in more specific themes which are supposed to cover the different features of the – temporal, spatial, social, … - context of the production of a discourse (such as an interview).

2. **Discourse generalities**. Theme that is deployed in a set of more specific themes covering what is called more technically the "signaletic information" or again the *paratextual* features of a produced discourse: title of the discourse, summary, key words, classification, copyright holder, etc. It is easy to see that this theme may be specified (if necessary and useful) by the concerned upper categories of the Dublin Core[58] enabling in this way the PCI ontology to play a "bibliotheconomical" function.

3. **Discourse participant**. Theme that, for the moment, is not developed in a sophisticated way and which recovers only the two or three main roles that are always implied in discourse production: the producer or author and the receiver or destine. It has to be stressed the fact that "author" or "producer" of a discourse is not necessarily the "uttering subject" what can be exemplified with the help of the example of the *quotation*: the author (of an interview, an exposé, a story, …) may quote somebody else for developing his purpose but it is the "somebody else" who is the uttering subject, i.e. the role that has the responsibility of what is uttered. This distinction has not been yet introduced in the PCI ontology but it will be in the next version.

---

[58] Cf. the DCMI metadata terms: http://dublincore.org/documents/dcmi-terms/#H2

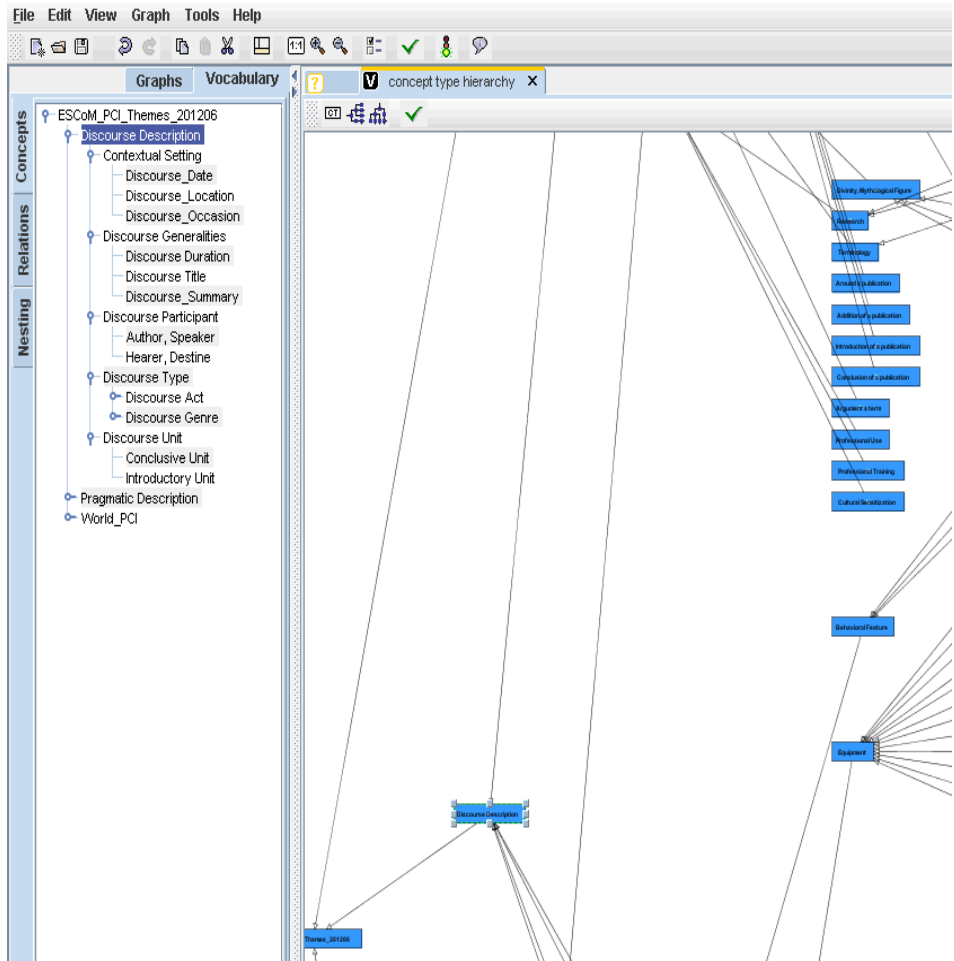

(*figure 7*: the narrative facet of the PCI ontology)

4. ***Discourse type***. This central theme (cf. figure 8) is deployed in two more specialised ones: the *Discourse* (or speech) *act* theme deploying itself a rather developed hierarchy of classes and types of speech acts as they are traditionally handled in speech act theory (Reinach, Austin, Searle, Vandervecken,…)[59] or in pragma-linguistic discourse analysis (Wunderlich, Maingueneau,…); the *Discourse genre* theme that recovers, for the moment, a small unstructured list of themes identifying specific genres such as *Summary, Interview, Chronology*, etc. The difference between these two themes – *Discourse Act* and *Discourse Genre* – is a distinction in terms of *compositionality*: a discourse genre is composed by one but habitually more than one discourse acts. For instance the discourse genre *Portrait* (of a social group, an animal, …) can be composed by the discourse act *Description* alone (in this case, we have a sort of a "pure" or prototypical portrait) or by a – syntagmatically – integrated set of discourse acts (for instance: a portrait of a social group composed by a description, a narration, an appreciation, etc.). Again, it's easy so see that the theme *Discourse* genre can become the upper category for a lot of controlled vocabularies related to Genres such as those presented and used by the

---

[59] Cf. Kent Bach foir a small comprehensive online introduction in speech act theory:
http://online.sfsu.edu/~kbach/spchacts.html

Library of Congress for classifying sound records[60]; of the cinematographic genres vocabulary as proposed in the instructions for classifying and indexing audiovisual and film resources by the FIAF (*Fédération Internationale des Archives du Film*)[61], etc.

5. *Discourse unit*. This them identifies generic and/or named parts or units composing a discourse genre or discourse act – units such as *Introduction*, *Conclusion*, *Development*, *Complication*, *Digression*, etc. It is very similar to what Louis Chammings from INA Recherche calls "narrative unit", i.e. a *functionally typified sequence in the development* of some genre such as, in Chammings' case, the *news* genre. This hierarchy has not yet been developed in the actually used 4th version of the PCI ontology. The fifth version, available at the end of 20007, will include this hierarchy which is particularly useful for decomposing (and reassembling) a specific discourse genre or act in identifying different typical and even critical steps or moments such as the introduction for presenting the portrait of a social group, the (dramatic) complication of it, digressive moments, etc.

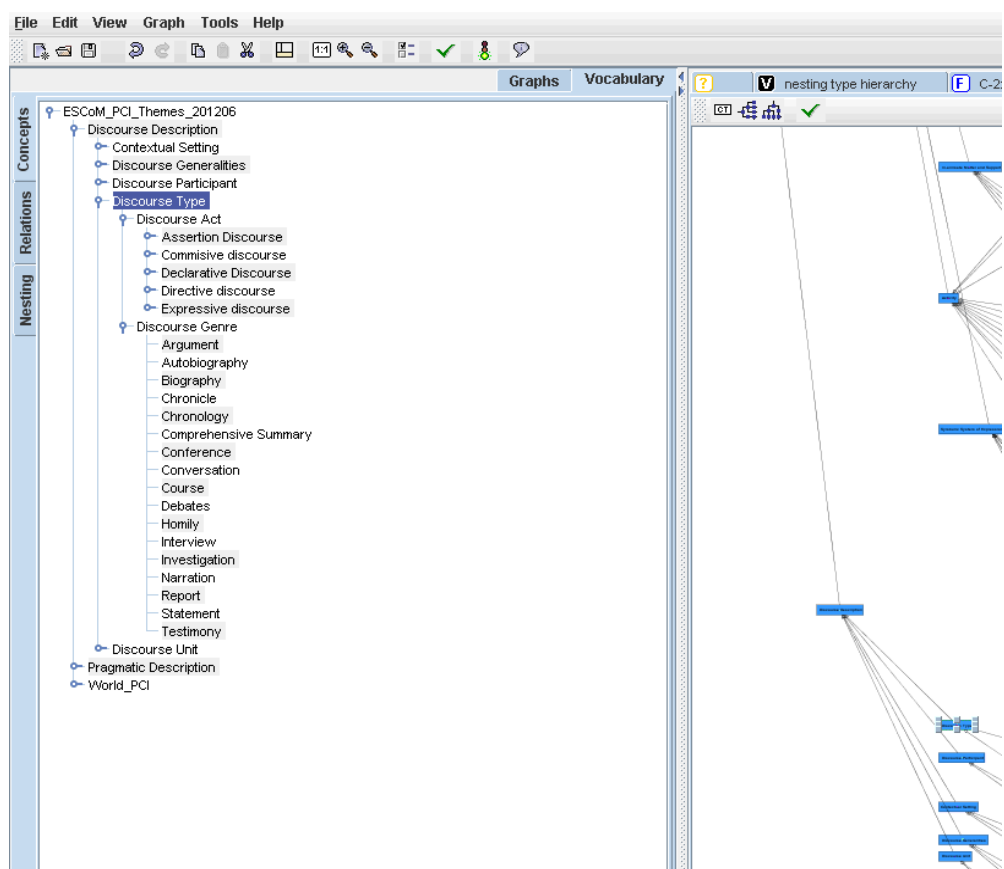

(*figure 8*: *Discourse Act* and *Discourse Genre* themes belonging to the *Discourse type* hierarchy)

It has to be noticed that the *Discourse Description* facet of the PCI ontology will be again developed in a more detailed and concise way in order to make it usable also for more fine grained discourse analysis including, for instance, the description of specific units (discourse units) composing a genre such as the argumentative genre used in one of the above mentioned highly innovative and creative CWI projects[62], the description of specific discursive or enunciation features such as the focussing, the highlighting, … of information, the use of metaphorical and analogical "images", etc.

---

[60] Cf. the recorded sound reference center of the LoC http://www.loc.gov/rr/record/gen.html
[61] Cf ; the online available instruction guides for the classification and indexing of cinematographic and audiovisual resources: http://www.fiafnet.org/fr/publications/fep_cataloguingRules.cfm
[62] Cf. note 11 concerning the very appealing and innovative CWI *Vox Populi project*

For the completion of the narrative facet *Discourse Description*, the TEI – especially the TEI facet "*Text Description*"[63] – will constitute a very valuable input as well as, naturally, the specialised literature in discourse description and linguistics of enunciation.

## *4.2/ The referential facet of the ontology: scenes form the PCI world as discourse topics*

The basic themes specialising the taxeme (upper category) *PCI_World* which constitutes the central referential facet of the PCI ontology are (cf. figure 8):

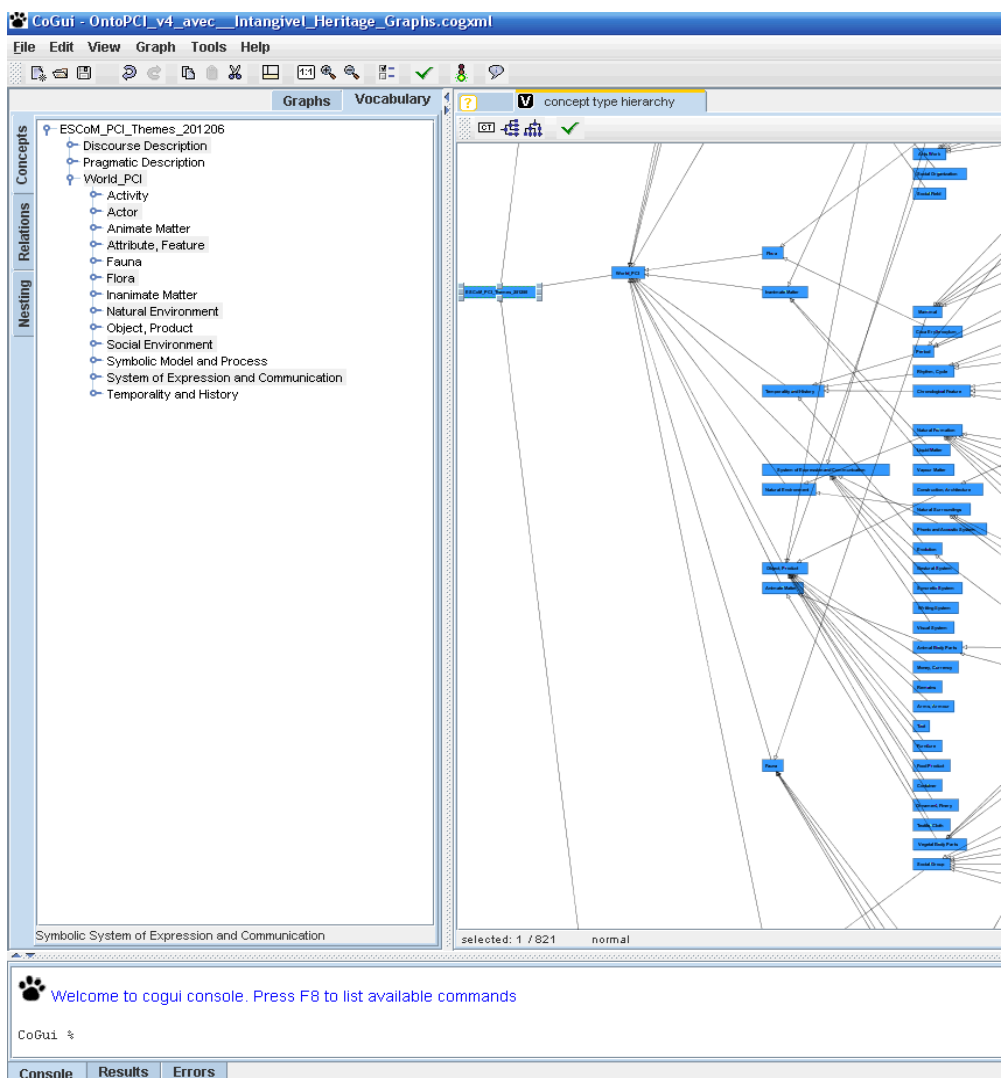

(*figure 8*: the upper categories, i.e. basic themes of the PCI World)

---

[63] Cf. TEI text description document:
http://www.hti.umich.edu/cgi/t/tei/tei-idx?type=extoc&byte=1964152
http://www.hti.umich.edu/cgi/t/tei/tei-idx?type=pointer&value=CCAHTD

1. ***Activity***. A theme that is deployed in an already rather developed hierarchy of specialised themes representing activities (doings, techniques, social practices, …) that characterise typically the "behaviour" of an actor (a person, an institution, even a non-real actor like a divinity, and – in our case: social groups such as minorities, aborigines, indigenous people, "ethnies", etc.). Figure 9 proposes a more explicit picture of the actually available hierarchy of themes by the means of which specific types of activities will be described. It is clear that this basic theme represents a central, trans-disciplinary category in social sciences which therefore needs certainly again a more focussed conceptual make up. It also has to be critically compared with different controlled vocabularies proposed for classifying social practices, professional activities, etc.

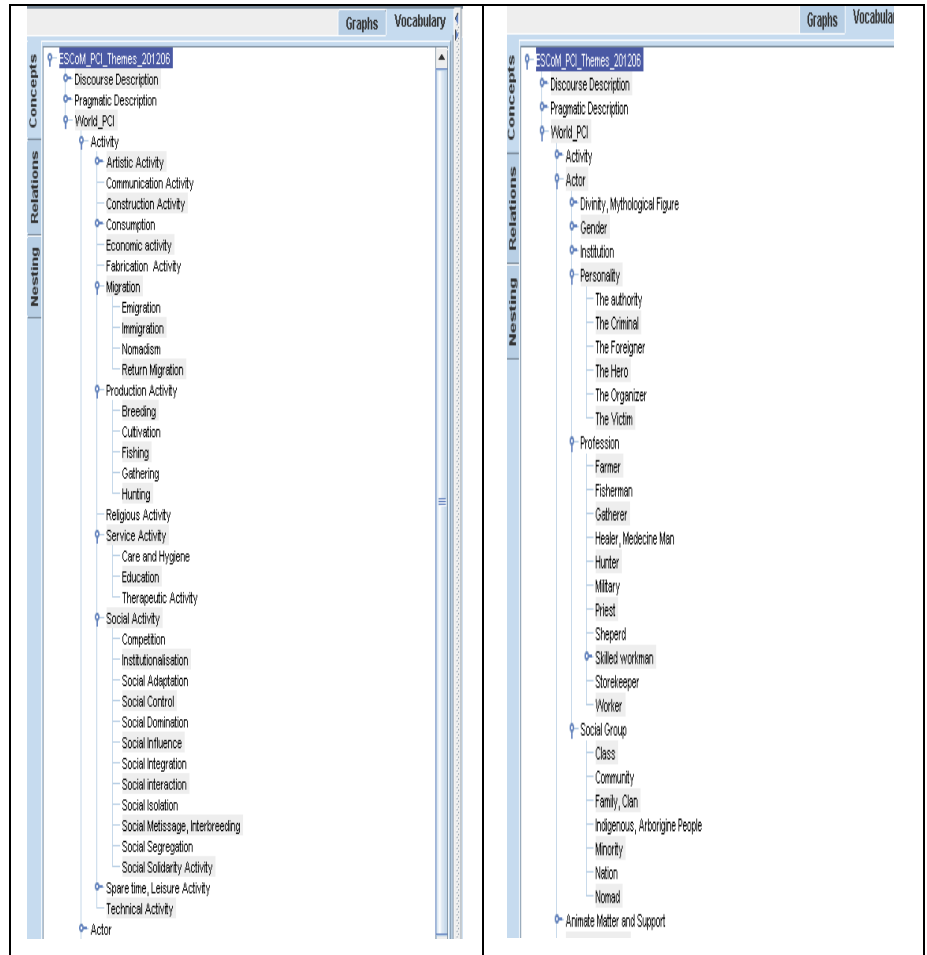

(*figure 9*: the actor and the social practice theme hierarchies)

2. ***Actor***. Like the *Activity* theme, also this theme recovers a set of central themes necessary for the description and indexing of the audiovisual corpus of the PCI web archive. Actors are – technically speaking *anthropomorphic roles*, i.e. "*human like roles*" or roles occupied by entities behaving like "humans". They constitute **one** of the **two** essential **role types** of a "life world" of human groups such as those that constitute the investigation object of the PCI audiovisual corpus; the **second** essential role type is represented by the *Object, Product* theme hierarchy that will be presented below. Concerning the *Actor* theme hierarchy, more specific roles have been identified in the audiovisual reference corpus among which, more particularly the two central roles *Minority* and *Indigenous People* which are specialisations of the *Social Group* theme. Figure 9 proposes an explicit summary of the actual state of the art of this hierarchy which also will certainly evolve in the next months given the fact that it represents one of the most difficult fields of investigation in social sciences.

3. *Animate Matter*. This basic theme covers a whole sophisticated set of more specialised themes which represents the *autonomous parts* of living beings or "complexes" classified in the two theme hierarchies *Fauna* and *Flora*. Autonomous parts are, for instance, the organs of an animal, the leaves of a plant, etc. Figure 10 proposes an explicit summary of the actual state of the art of this hierarchy which will (monotonously) evolve following the empirical needs for describing the audiovisual corpus of the PCI pilot.

4. *Attribute and Feature*. This basic theme covers the important category of *non-autonomous* entities of living beings, inanimate entities as well as abstract (conceptual) entities. This indeed as central as complex category, is for the moment only developed in its basic articulations. The empirical study of the PCI corpus will show what has to be done in the coming months for updating the *Attribute and Feature* theme hierarchy. Figure 10 shows the actual state of the art of this hierarchy.

5. *Symbolic Model and Process*. This basic theme circumscribes the domain of what is called the **cultural**, i.e. the **cognitive** and **axiological** (value) framework of the "life world" of an actor, the technical skills and knowledge for performing activities, beliefs and knowledge concerning the surrounding world, rules and traditions governing the doing and behaviour or again the canons of truth, beauty, morality, pleasure, utility, etc. Figure 12 shows the actual state of art of *Symbolic Model and Process* theme hierarchy.

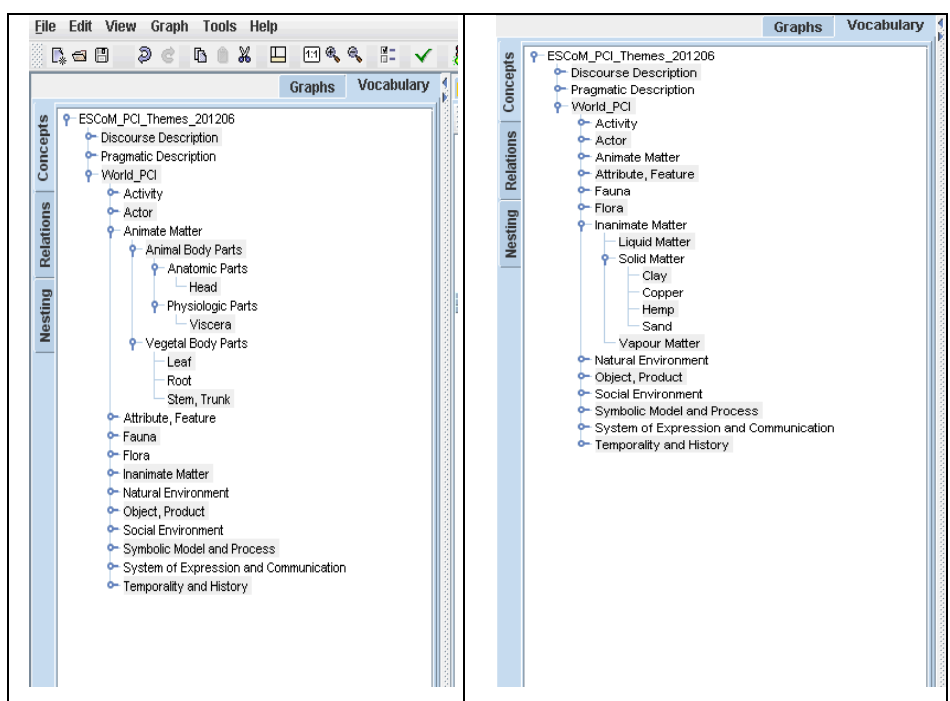

(*figure 10*: animate and inanimate matter and support theme hierarchy)

6. *Fauna*. This basic theme develops the "animal kingdom" - the world of animals, human species included. It will be (monotonously) developed following the empirical description and indexing needs of the audiovisual corpus of the PCI pilot. One of the inspirational sources for the elaboration of this hierarchy of themes is the beautiful *Tree of Life Web* project of the NSF and the University of Arizona[64].

7. *Flora*. This basic theme recovers the "vegetable kingdom" – the world of plants. It will be (monotonously) developed following the empirical description and indexing needs of the audiovisual corpus of the PCI pilot. One of the inspirational sources is, once more again, the beautiful *Tree of Life Web* project of the NSF and the University of Arizona[65].

8. *Inanimate Matter*. This theme recovers the world of (solid, liquid and vapour) matter and will be (monotonously) developed following the empirical description and indexing needs of the audiovisual corpus of the PCI pilot. Figure 11 shows the yet very cumbersome hierarchy dedicated to this referential domain.

9. *Natural Environment*. This theme recovers the landscape and the natural formation as well as the physical ecology of the "life world" of a social actor such as a social group, a minority, etc. Figure 11 shows the actual state of the art of this hierarchy which, once more again, will be developed according to the empirical description and indexing needs of the audiovisual corpus of the PCI pilot.

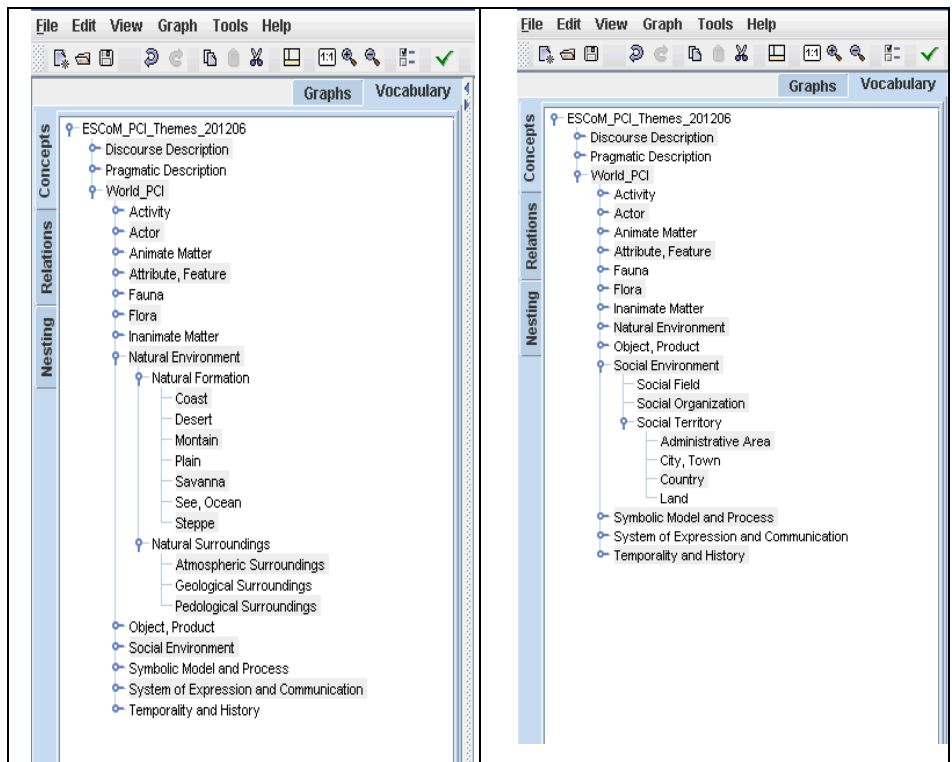

(*figure 11*: the natural and social environment theme hierarchies)

10. *Object, Product*. This central basic theme recovers another, second *main type of roles* (the first one is the social actor role represented by the *Actor theme* hierarchy); it recovers, in other words, more specific roles such the tool & instrument role, the equipment role, the decoration role, the container role, the information and knowledge support role, etc. that *any* entity (animate and inanimate matter, animal, plant, …) may occupy in an activity or social practice or, more generally speaking, in the world of a given minority or social group. Figure 12 shows the actual state of the art of the *Object, Product* theme hierarchy.

---

[64] cf. the web site of the Tree of life web project: http://tolweb.org/tree/phylogeny.html
[65] cf. the web site of the Tree of life web project: http://tolweb.org/tree/phylogeny.html

11. ***Social Environment***. This basic theme recovers – a not yet very developed – hierarchy of specialised themes that are concerned by the description of the social context in which a given social group is living. This social context is differentiable in an "internal" and an "external" one. The "internal" social context means that a social group possesses some kind of organisation, represented by the theme *Social Organisation* and exemplified, for instance, by the fact that a social group possesses a clan like structure, a caste like structure, a feudal structure, a "tribal structure", etc. The "external" social context means that a social group is living in a social (political, economical,…) environment represented by the theme *Social Territory* which may be specialised as an administrative one, a political one, a linguistic one, a religious one, etc. Figure 11 shows the actual state of the art of this theme hierarchy.

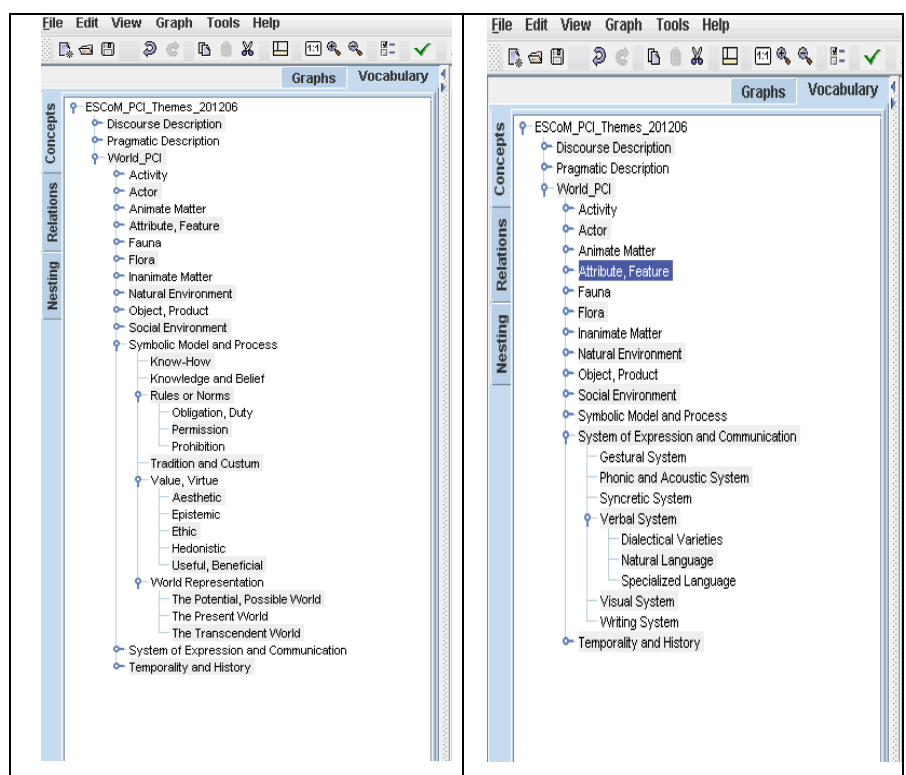

(*figure 12*: the cultural reference & form theme hierarchy and the symbolic system of expression & communication theme hierarchy)

12. ***System of Expression and Communication***. This basic theme of the referential taxeme covers the domain of language or languages belonging to the world of a social group (a minority,…) to be described. In version 4 of the PCI ontology, it is only developed (cf. figure 12) with respect to the main types of semiotic systems used by any actor for expressing an idea, an information, a knowledge, etc. and for communicating it. But the Verbal System theme is already developed in another ESCoM ontology dealing with a corpus on linguistic and cultural diversity[66] in taking into account already existing linguistic ontologies such as GOLD[67] or controlled vocabularies such as *SIL's glossary of language study*[68] or again the *Ethnologue language name index*[69] of more than

---

[66] This DLC ontology ("DLC" stands for *Diversité Linguistique et Culturelle*) is used for the description and indexing of an other audiovisual corpus which can be accessed via the DLC web portal: http://semioweb.msh-paris.fr/corpus/dlc/FR/
[67] GOLD: General Ontology for Linguistic Description: http://www.linguistics-ontology.org/gold.html
[68] SIL International (formerly known as Summer Institute of Linguistics): http://www.sil.org/linguistics/glossaryoflinguisticterms/contents.htm
[69] Cf. http://www.ethnologue.com/language_index.asp

7000 primary language names. In the coming months, it will be decided of how to use this part within the descriptive and indexing work of the PCI corpus (if needed): either as an imported part from the DLC ontology or as a common part for both ontologies (as well as for other ones …).

13. ***Temporality and History***. This basic theme deals with the temporal and historical setting of the life world of a given social group – a minority or indigenous people. Figure 13 shows that the basic articulation of this hierarchy takes into account 4 temporal characteristics: the simple *Chronology* which nevertheless is culturally dependent on a sort of theory adopted by a group or social groups for quantifying the temporal flux; the *Period* for localising the life world of a social group or some part of it – characteristics that, once more again, may vary with respect to different cultural models; the *Rhythm &* (periodical) *Cycle* following which social groups organise their (personal, collective, ritual, political, …) agenda; and finally the very important *Evolution* characteristics concerning the (social, demographic, natural, …) changes within the life world of a social group (or a part of this life world).

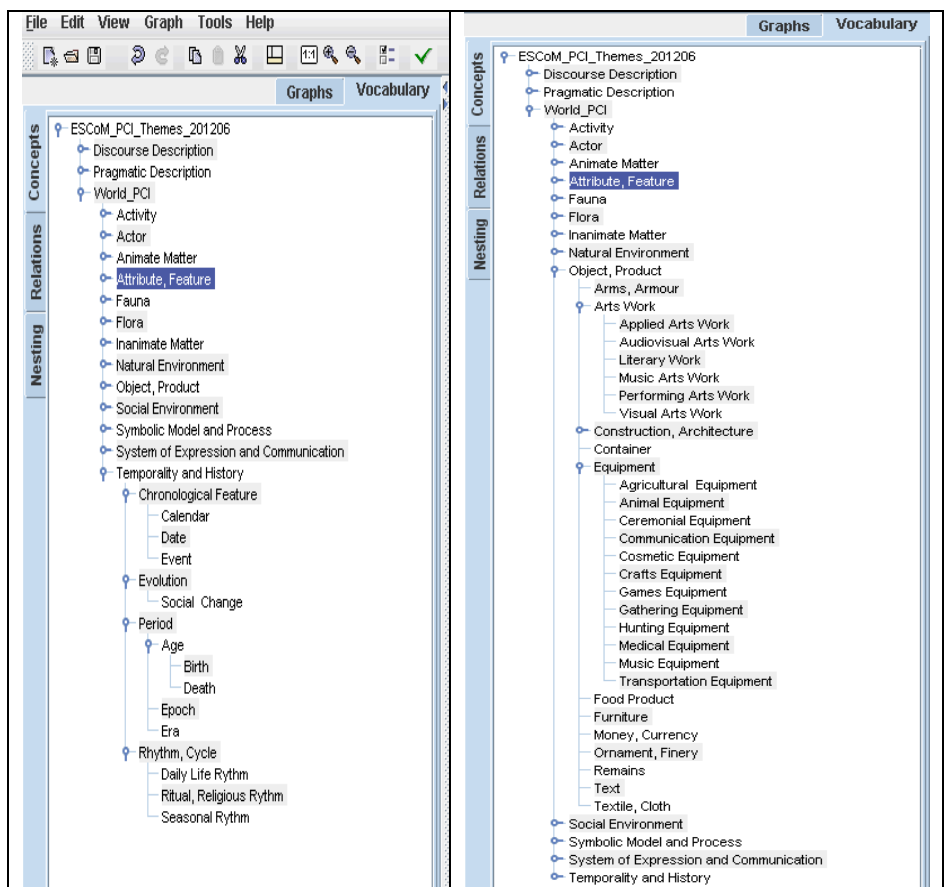

(*figure 13*: the temporality& history theme hierarchy and the object, product theme hierarchy)

# 5/ The relational part of the PCI ontology

The meta-term "relation" is strictly similar with the meta-term "property" in the CIDOC Conceptual Reference Model (version 4)[70]. But contrarily to the not-structured list of more than 100 properties populating the CRM of CIDOC, the relations in the PCI ontology are categorised and, this, following linguistic, cognitive and semiotic researches and "standards" such as linguistic case grammars, lexical relational models, rhetorical structure theories, pragmatic and argumentation theory, narrative semiotics, cognitive modelling of belief, discourse and action, etc.

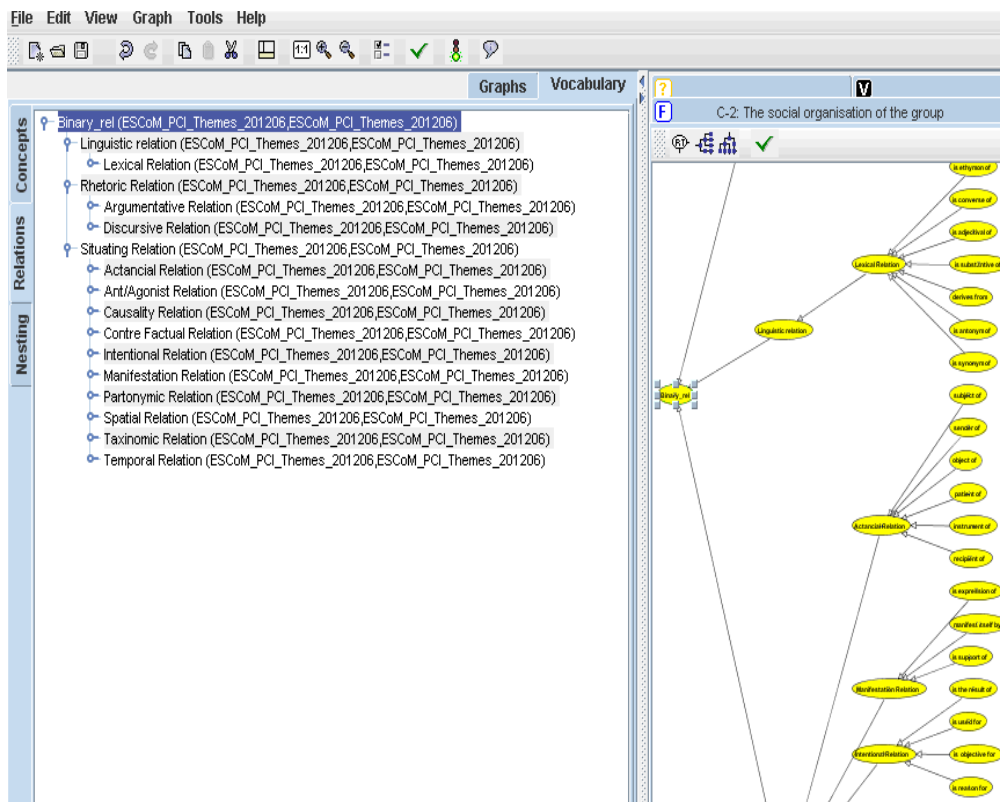

(*figure 14*: the relational theme hierarchy of the PCI ontology)

The PCI relational ontology is divided actually in three upper categories or taxemes (figure 14) by the means of which thematic configurations (i.e. conceptual graphs) representing discourse topics relative to the cultural heritage of minorities and social groups will be specified and elaborated:
1/ The *Situating Relation* taxeme;
2/ The *Narrative Relation* taxeme;
3/ The *Linguistic Relation* taxeme.

---

[70] Cf. the online version of CIDOC's CRM, version 4: http://cidoc.ics.forth.gr/docs/cidoc_crm_version_4.0.doc

## 5.1/ The Situation Relation Taxeme

This taxeme recovers a variety of basic types of relations by the means of which "scenes" of the life world of a social group (a minority, an indigenous people) could be described and indexed – *scenes* relating, for instance, technical or artistic activities of a social group, scenes relating the belief and the customs of a social group, scenes relating the place of a social group within a socially determined territory, scenes relating the internal social organisation of a group, scenes relating the fauna and the flora in the life world of a social group, scenes relating the demographic and social evolution of a group, and so on.

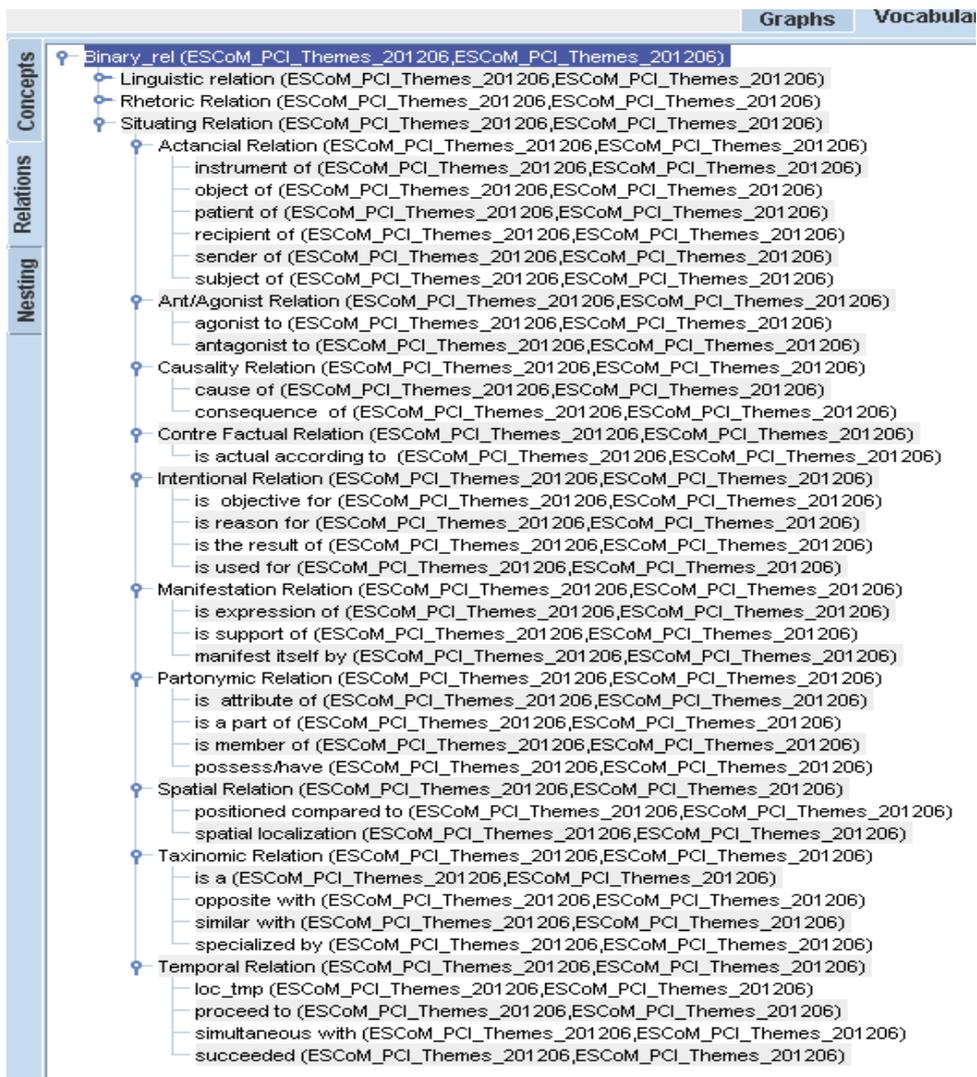

(*figure 15*: the situating relations hierarchy of the PCI ontology)

The actually identified and more or less developed relational theme hierarchies composing the *Situating Relation* taxeme (cf. figure 15) are classified in the ten following basic themes:

1. The ***Actantial*** or casual ***relational theme hierarchy*** that recovers basic action roles participating in an activity, doing, social practice, etc – roles such as the subject (the instigator) of an activity, the patient (the beneficiary) of an activity, the object of it, etc;

2. The *Ant/Agonist relational theme hierarchy* that recovers the basic roles in concurrent and opposed actions such as conflicts, battles but also concurrent actions for achieving a common or opposed goals;

3. The *Causality relational theme hierarchy* that refers to the basic causal relationships between two more events, actions, etc.

4. The *Counterfactual relational theme hierarchy* that recovers the basic relationships between two different world states in the life world (of a social group) such as an actual state (of, let's say social organisation, place in a social hierarchy, …) and an imaginary one (of social organisation, place in a social hierarchy, etc.);

5. The *Intentional relational theme hierarchy* that recovers the basic relationships necessary for representing projects, objectives, motives, etc. in the life world of a social group such as for instance the motives of a ritual activity, the objectives of it, the means and resources for accomplishing it, etc.;

6. The *Manifestation relational theme hierarchy* that recovers the relationships between a symbolic form and its (physical) support or again a project, a plan, a model and its realised, perceptible form, etc.;

7. The *Partonymic relational theme hierarchy* that recovers the compositional relationship between different entities forming a "complexion" such as, for instance, the natural formation "island" which constitutes the natural surrounding of a social group and which is – so to speak – composed by a fauna, a flora, inanimate matter and animate matter as well as typical objects (testifying the skills and traditions of a social actor), and so on;

8. The *Spatial relational theme hierarchy* (not yet developed) that recovers the spatial location and direction relationships between entities of the life world of a social group such as the flora located in the middle of an island or on its costal borders, in the north or in the south, etc.

9. The *Temporal relational theme hierarchy* (also not yet very developed) that recovers the temporal positioning of entities of the life world of a social group;

10. The *Taxonomic relational theme hierarchy* that allows to redefine taxonomic relationships with respect to the specific culture of a given life world (that means that a taxonomic relationship, for instance, between two vegetal species defined in the PCI hierarchy may be "redefined" with respect to the specific ethno-botanic culture of an indigenous group).

## 5.2/ The Narrative Relation Taxeme

The *Narrative Relation* taxeme recovers all relationships between any theme of the PCI ontology with a theme from the Discourse Description theme hierarchy. This relational taxeme is divided in two basic relational themes (figure16):

1. The *Discourse relational theme hierarchy* which defines all specific relationships between the different discourse units composing the discourse produced by its author (in our case: a researcher, teacher, etc.);

2. The *Rhetorical relational theme hierarchy* which defines the argumentative relationships between discourse acts produced in order to transmit a specific message. For instance, the description of the rituals of a social group may be intended to exemplify a specific ethnological theory of rites; the narration of care taking practices of children may be intended to explain or again to back up a specific hypothesis, etc.

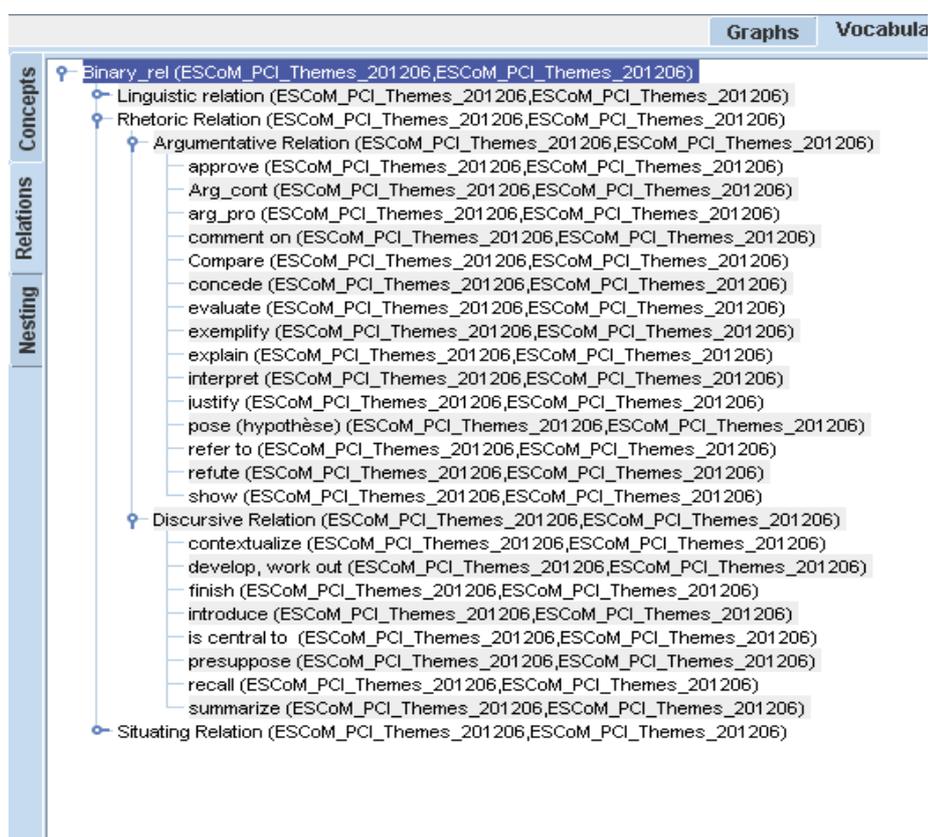

(*figure 16*: the narrative relations hierarchy of the PCI ontology)

The Discourse relational theme hierarchy will be developed in a more detailed way for the next version of the PCI ontology. Its development depends strongly on abetter definition of the *Discourse Unit* theme hierarchy (belong to the narrative facet of the PCI ontology, cf. chapter 4.1).

The *Rhetorical relational* theme hierarchy also will be more explicitly developed, especially with reference to the *Rhetorical Structure Theory*[71] of W.C. Mann and S. Thompson.

## 5.3/ The Linguistic Relation Taxeme

The *Linguistic Relation* taxeme (figure 17) has so far only be developed with respect to the empirical needs of description and indexing of the audiovisual corpus composing one of the pilots of the Logos project. Concretely speaking, lexicological relationships positioning terms or terminological syntagma have been defined for describing and indexing "indigenous vocabularies" This means that, in its actual form, the Linguistic Relation taxeme only recovers a small and very specific part of linguistic relations, i.e. relationships between linguistic entities such as morphemes, lexical items, noun or verbal phrases, etc. If there is a specific need, this taxeme will be deployed more ambitiously following linguistic "standards" such as the GOLD ontology or SIL's contribution to language studies[72].

---

[71] Cf. the RST web site : http://www.sfu.ca/rst/
[72] Cf notes 57, 58, 59, 60

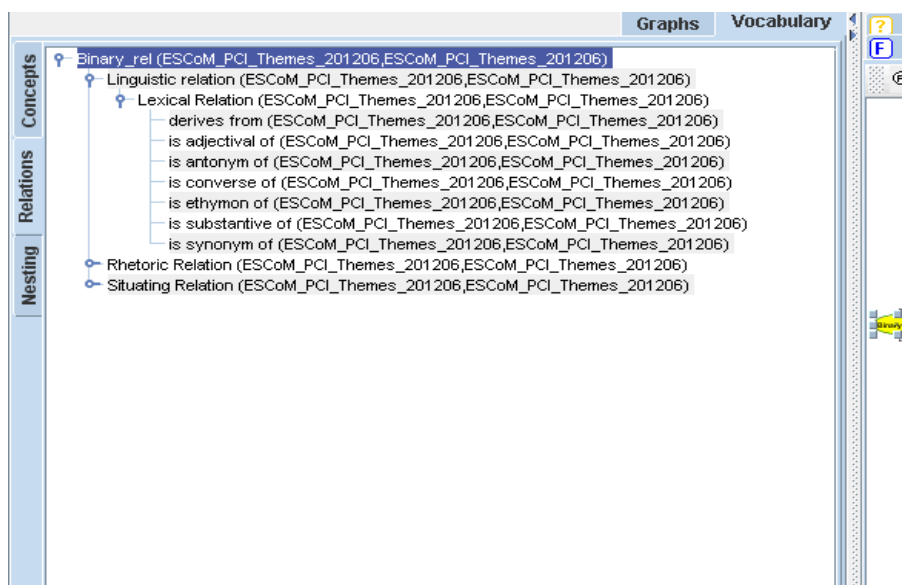

(*figure 17*: the linguistic/lexical relations hierarchy of the PCI ontology)

# 6/ Conceptual graphs and thematic configurations

## 6.1/ The main types of thematic configurations

As already developed in the first chapter of this paper, conceptual graphs are, simply speaking, composed of a set of *themes* that are connected each other via *relations* – thematic relations or again relational themes. Themes connected in this way are called *thematic configurations*.

One central type of thematic configurations is representing *scenes* of the domain of reference, the domain of knowledge, *viz*. in our case, scenes dealing with the cultural heritage (not in the restrictive museological like sense but in the broad sense used in social sciences) of social minorities and so-called indigenous populations. Such scenes are for instance, as shown in figure 19, scenes concerning specific types of activities characterising the daily life of a minority or an indigenous population, scenes characterizing the natural and the social environment of a minority or an indigenous population, scenes dealing especially with the artistic achievements of a studied social group, etc. This type of thematic configurations is called *topical configurations* or again *topics*.

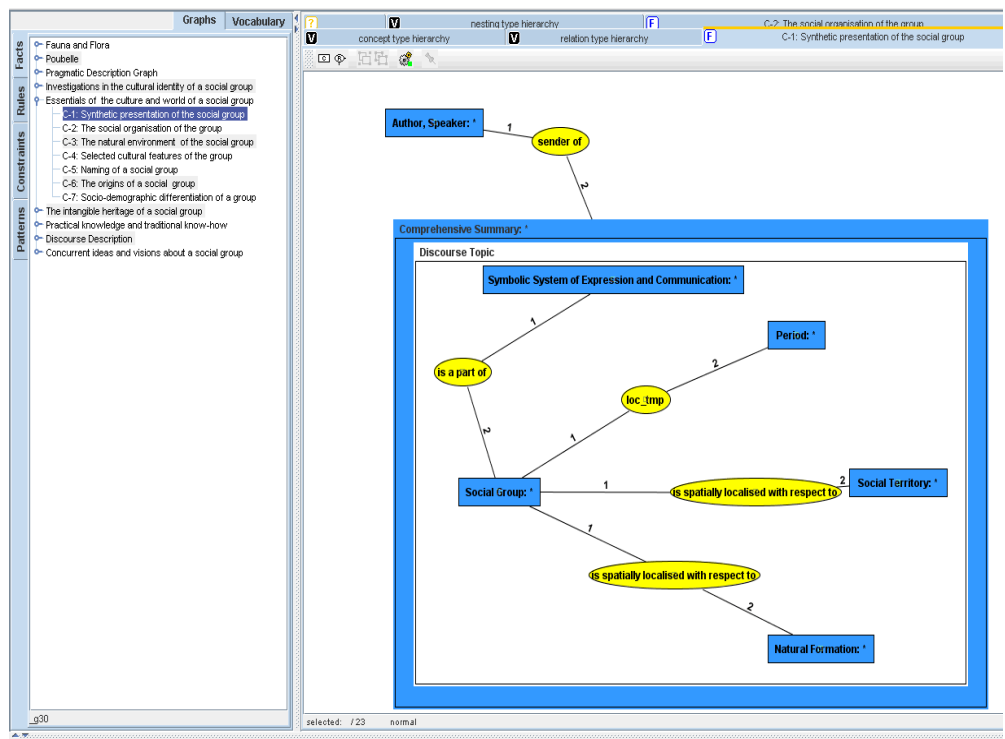

(*figure 18*: example of a conceptual graph representing a discourse topic)

A second important and central type of thematic configurations represent the discourse(s) by the means of which someone (the speaker, the author) "speaks" about a minority, an indigenous population, develops the one or the other of the scenes characterising the life world of the referred social group, its daily life, etc. This type of thematic configurations is called *narrative* or again *narrative configuration* (in the sense of A.J. Greimas and his narrative grammar).

A third important type of thematic configurations represents the context of use or again the context of possible exploitations of a produced discourse and its message with respect to given parameters such as, for instance:
- the cultural and cognitive profile of the "user" of a discourse,
- the social settings of the possible reuses of a produced discourse,
- and so on.

This type of configuration is called – *faut de mieux* – *pragmatic configuration*. It has to be noticed that, in the context of the Logos project, pragmatic configurations will not be used because they are a part of the authoring studio of the Logos project.

In taking the point of view of text or discourse semiotics, there are again other types of thematic configurations such as especially *configurations* representing the *expression and organisation modalities of a text* (*lato sensu*, i.e. including not only "written" one but also visual and audiovisual ones)[73]. This – and other – types of configurations will not be developed in the context of the Logos project.

It has to be noticed that in the context of the PCI pilot, topics are always ***discourse topics***. This means that we always have to do with at least ***two graphs*** whereas the one – representing a topic – is embedded in the other – representing the discourse by the means of which a speaker frames and develops the topic. The topic is the *referent* of the *discourse* and the topic is always determined by the *point of view* of the speaker – the researcher, the teacher, etc.

The conceptual graph representing the narrative configuration is therefore also called the "outer graph" and the conceptual graph representing the topical configuration, the "inner graph" (figure 18).

## *6.2/ Identified discourse topics*

The PCI audiovisual corpus will be described and indexed with respect to a list of discourse topics which have been already identified in *Step 1* (of the elaboration of the ontology) consisting mainly in the empirical work on the available corpus[74]. Actually 4 principal groups of discourse topics are defined (cf. figure 19):

1. *Essentials of the culture and life world of a social group:* discourse topics comprising in a summarized form general information about the identity of a social group, its social organisation, its culture, its natural and social environment, its language, etc.

2. *The intangible heritage of a social group*: discourse topics developing as systematically as possible different aspects of the language of a social group, the artistic works of it as well as well as its religious references (this group of discourse topics corresponds rather strictly to UNESCO's definition of "intangible heritage");

3. *The practical knowledge and traditional know how of a social group*: discourse topics developing information about the technical and practical activities characterising the daily life of a social group, the practical intelligence necessary for performing these activities, the tools and equipments used for performing these activities, etc.;

4. *Investigations in the cultural identity of a social group*: discourse topics developing information about the "cultural self" of a social group, its relationships with the social other, its position within a social environment, etc.

---

[73] Cf. the already quoted CWI project "*Towards ontology-driven discourse: from semantic graphs to multimedia presentations*" is based on three "ontologies" (footnote 10)

[74] cf. chapter 3 for more information

Other groups of discourse topics will be specified again, especially related to the history and the evolution of a social group, to its temporal agendas or again to its material culture. But such as, the above four quoted groups of discourse topics already represent a significant coverage of the content provided by the audiovisual corpus of the PCI pilote.

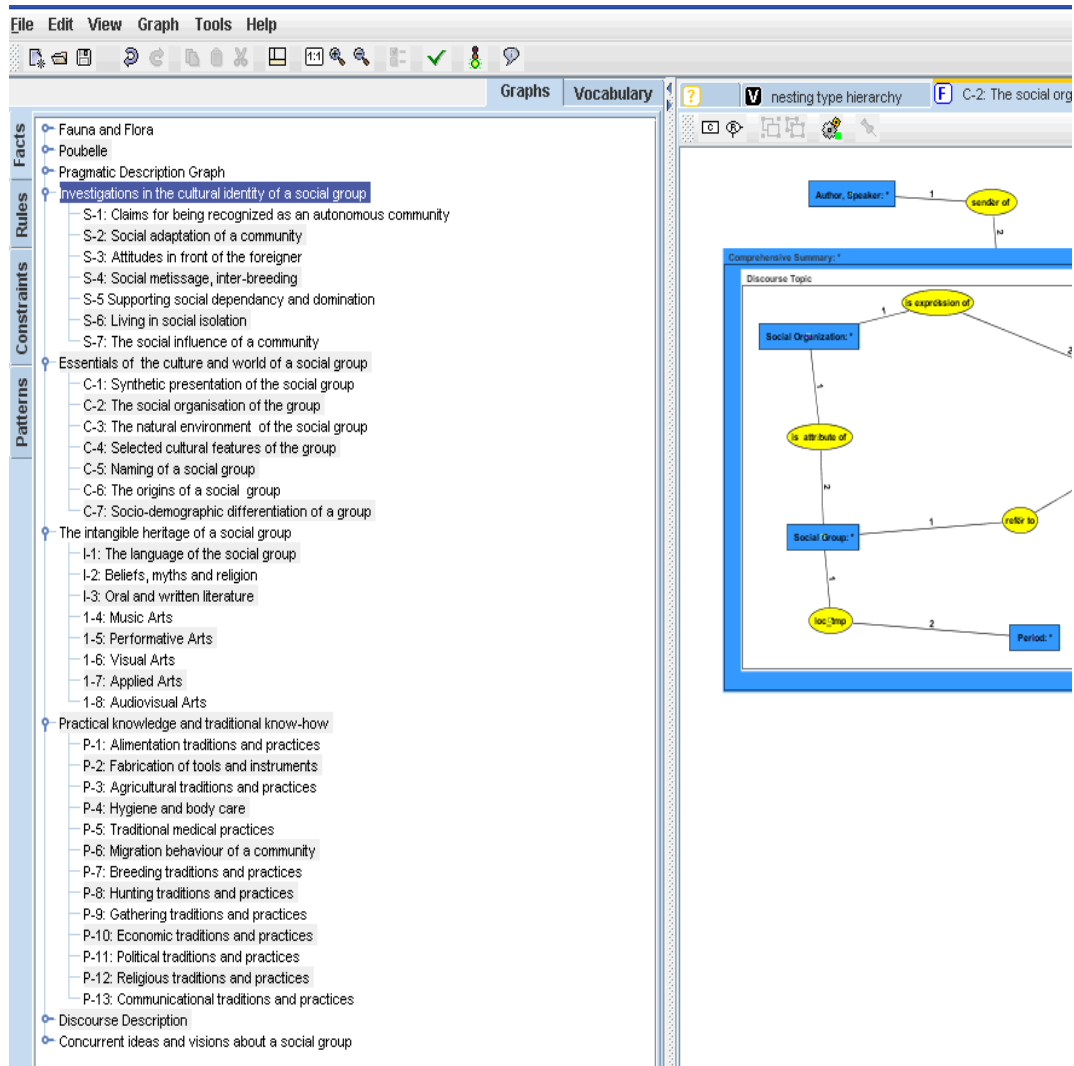

(*figure 19*: list of conceptual graph templates for indexing the PCI audiovisual corpus)